# Machine-learning-guided molecular dynamics simulations of point defect evolution in $\beta$-$Ga_2O_3$ during ion implantation and annealing

Huawen Li[1], Mengzhi Yan[2], Zongwei Xu[1,*], Junlei Zhao[3], Jiale Wang[1]

[1]*State Key Laboratory of Precision Measuring Technology and Instruments, Laboratory of Micro/Nano Manufacturing Technology (MNMT), Tianjin University, Tianjin 300072, China*

[2]*Centre of Micro/Nano Manufacturing Technology (MNMT-Dublin), School of Mechanical and Materials Engineering, University College Dublin, Belfield, Dublin 4, D04 V1W8, Ireland*

[3]*The Hong Kong Microelectronics Research and Development Institute, Hong Kong, 999077, China*

[*]Corresponding Author: zongweixu@tju.edu.cn

**ABSTRACT**

In beta-gallium oxide ($\beta$-$Ga_2O_3$), Ga-ion implantation and subsequent annealing induce abundant point defects, including interstitials and vacancies. To overcome the limitations of conventional Wigner-Seitz (WS) defect analysis, a defect identification algorithm based on similarity matching and spatial clustering via density-based spatial clustering of applications with noise (DBSCAN) is developed specifically for $\beta$-$Ga_2O_3$. This algorithm accurately distinguishes lattice atoms from point defects under thermal perturbations and at high defect concentrations, and further identifies eight configurations of Ga interstitials ($Ga_{ia}$ to $Ga_{ih}$) by analyzing atomic coordination environments. Simulations comparing Stopping and Range of Ions in Matter (SRIM) and molecular dynamics (MD) data, along with single-ion implantation analysis, highlight the significance of electronic stopping effects: neglecting electronic stopping leads to overestimated ion range, defect concentration, and temperature rise. Across five implantation fluences (1 to 5 × $10^{14}$ $cm^{-2}$) and corresponding annealing processes, 1373 K is identified as the optimal recovery temperature. Multiscale analyses based on hydrostatic stress, partial radial distribution function (PRDF), and defect concentration reveal the evolution and spatial distribution of point defects. The results demonstrate that Ga interstitials ($Ga_i$) tend to occupy tetrahedral and octahedral interstitial sites, and the accumulation and recombination of point defects drive a defect-mediated phase transition from *β*- to *γ*-$Ga_2O_3$. With increasing fluences followed by annealing, *β*-phase recovery declines while *γ*-phase transformation rises, leading to an irreversible phase transition. In contrast, the migration of oxygen interstitials ($O_i$) is more sensitive to annealing temperature, and appropriate annealing temperature significantly enhances recrystallization of the O-sublattice.

## 1. Introduction

Gallium oxide ($Ga_2O_3$) is an emerging ultrawide-bandgap semiconductor whose most notable characteristics are an exceptionally large bandgap of approximately 4.8 eV and a high theoretical breakdown electric field of up to 8 MV/cm. These properties render it a promising candidate for high-power electronic devices and solar-blind ultraviolet photodetectors [1,2]. Among its five polymorphs, beta-gallium oxide (*β*-$Ga_2O_3$) is the thermodynamically most stable phase [3] and exhibits superior irradiation tolerance. This resilience is attributed to its ability to form metastable crystalline gamma phase (*γ*-$Ga_2O_3$) upon ion implantation. The associated structural phase transition accommodates lattice damage rather than accumulating disorder, thereby effectively suppressing amorphization [4–6]. Ion implantation is a key technology for tailoring the electrical properties of *β*-$Ga_2O_3$, enabling precise doping required for device applications. Notably, a growing number of studies using atomically-resolved scanning transmission electron microscopy (STEM) have experimentally confirmed that ion implantation induces a phase transformation from the *β*-phase to the *γ*-phase in *β*-$Ga_2O_3$ [4–9].

Within *β*-$Ga_2O_3$, a variety of complex defect types exist. Johnson *et al.* [10] were the first to directly observe point defect complexes exhibiting a split configuration using STEM. Theoretically, numerous first-principles calculation studies have confirmed the existence of multiple defect complexes [11,12]. These split-vacancies or split-interstitials significantly reduce their formation energy by sharing lattice sites and act as compensation centers, constituting key factors influencing material diffusion behavior, irradiation response, and electrical properties [13,14]. They are directly linked to the irradiation tolerance, dopant diffusion, and device degradation behavior of *β*-$Ga_2O_3$ [15,16].

Recently, there has been a rapidly growing interest in classical molecular dynamics (MD) simulations of *β*-$Ga_2O_3$, with a focus on the threshold displacement energy (TDE) of primary knock-on atoms (PKA) to elucidate defect diffusion mechanisms [17–19], while relatively few studies have been reported on ion implantation and annealing under high-fluence conditions. To unveil the dynamic evolution mechanisms of point defects at the atomic scale, precise defect identification serves as a fundamental prerequisite. The Wigner-Seitz (WS) method, widely employed for point defect analysis in MD simulations, relies on spatial partitioning via Voronoi polyhedra. However, as noted by Yan *et al.* [20], the WS method exhibits limitations when applied to low-symmetry crystal structures such as $\beta$-$Ga_2O_3$. Furthermore, lattice thermal vibrations and the disturbance of the local atomic environment by high-concentration defects during ion implantation constitute the primary sources of error in defect identification.

In this study, a defect identification algorithm based on density-based spatial clustering of applications with noise (DBSCAN) clustering and similarity matching [21,22] was developed, specifically designed for the complex defect environment in $\beta$-$Ga_2O_3$. This algorithm effectively distinguishes point defects (interstitial atoms and vacancies) from lattice atoms under conditions of thermal perturbation and high defect concentration through dimensionality reduction projection and pattern recognition, successfully identifying eight types of Ga interstitial sites (from $Ga_{ia}$ to $Ga_{ih}$). Using MD simulations with a 2 keV Ga-ion source, the influence of electronic stopping power on ion implantation range, temperature, and defect concentration was preliminarily investigated. By analyzing five implantation fluences (1 to 5 × $10^{14}$ $cm^{-2}$) and utilizing metrics such as hydrostatic stress, partial radial distribution function (PRDF), and defect morphology, the evolution mechanisms of various point defects during ion implantation and annealing were systematically elucidated. The results reveal a key mechanism wherein the migration of $Ga_i$ into the eight interstitial sites induces the phase transformation from the $\beta$-phase to the metastable $\gamma$-phase. These atomic-scale insights provide a foundation for a deeper understanding of irradiation damage and recovery behavior in $\beta$-$Ga_2O_3$, while also offering a theoretical framework for optimizing ion implantation processes in power electronic devices.

## 2. Methods

### 2.1 Simulation Parameters Setup

All MD simulations in this work were performed using the Large-scale Atomic/Molecular Massively Parallel Simulator (LAMMPS) [23]. Visualization and post-processing were conducted with Open Visualization Tool (OVITO) [24] and Python. As illustrated in Fig. 1, the initial simulation model had dimensions of 9.6856

× 9.8743 × 18.814 $nm^3$ and contained 163,840 atoms, with the *z*-axis oriented along the [001] direction and the *y*-axis along the [010] direction (a detailed schematic of the crystallographic orientations is provided in the supplementary material Figure S2). The model was partitioned into three distinct regions: Boundary, Thermostat, and Newtonian layers [25]. The atoms in the Boundary layer at the bottom of the model were fixed in place to mimic a bulk substrate. The Thermostat layer utilized the Berendsen [26] thermostat to efficiently dissipate heat generated during implantation, while atoms in the Newtonian layer evolved according to Newton's equations of motion under the microcanonical ensemble (*NVE*). A 6 × 6 $nm^2$ area was designated as the implantation zone (the area of this region was utilized to calculate the ion fluence), with Ga ions being randomly generated 20 Å above this region. Ga ions were then injected toward the (001) surface with an incident energy of 2 keV. To minimize channeling effects, the incident beam was tilted by 7° [27] away from the (001) plane normal towards the [0-10] direction.

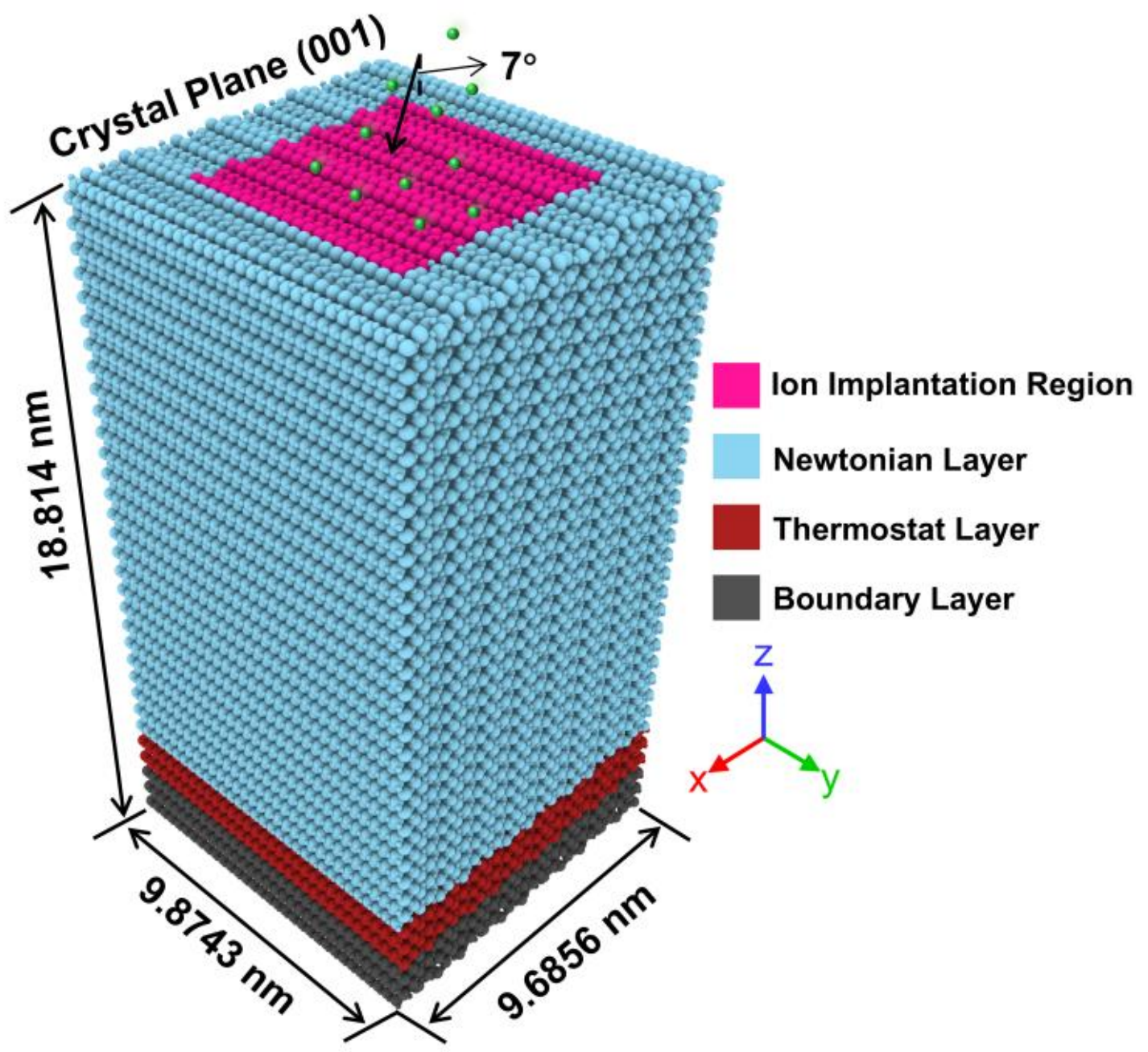

**Fig. 1.** Atomic model for Ga ion implantation into $\beta$-$Ga_2O_3$, containing 163,840 atoms with fix bottom Boundary (*z* in 0–1.5 nm), Thermostat (*z* in 1.5–3 nm), and Newtonian (*z* > 3 nm) layers along the *z*-axis. The ion implantation is in the pink square region of the size of 6 nm × 6 nm

The ion implantation and annealing simulations were conducted using the machine-learned tabGAP interatomic potential developed by Zhao *et al.* [28], which has been demonstrated to accurately and efficiently describe atomic interactions in $Ga_2O_3$ [4,17,29]. Prior to ion implantation, the model first underwent energy minimization followed by a 10 ps isothermal-isobaric ensemble (*NPT*) relaxation under 0 bar and 293K. During the implantation stage, the electronic stopping power and adaptive time steps were incorporated to enhance simulation accuracy [30–33]. Owing to the low thermal conductivity of $\beta$-$Ga_2O_3$, a relaxation period of 200 ps was applied after each ion impact to allow the system to dissipate the thermal-spike [34] and return to 293 K before introducing the subsequent ion. During the annealing stage, the entire system was subjected to heat treatment under the *NPT* ensemble using a Nosé-Hoover thermostat [35], with pressure control applied along the *x*- and *y*-directions. The temperature was increased from 293 K to the target annealing temperature (the reported temperatures refer to the global system temperature. Due to the inclusion of stationary atoms in the Boundary layers, which contribute to the thermal averaging, the actual temperature of the Newtonian layer is approximately 120 K higher than the prescribed system temperature) at a rate of 0.02 K/fs [36], followed by a 1 ns isothermal hold, and then cooled back to 293 K at the same rate. A final 5 ps relaxation was performed under the canonical ensemble (*NVT*). Simulations were carried out for five different implantation fluences, each followed by an annealing process at 1373 K. The corresponding parameters are summarized in Table 1.

**Table 1**

Simulation parameters of ion implantation and annealing.

| Condition | Parameters |
| --- | --- |
| Work material | $\beta$-$Ga_2O_3$ (001) |
| Boundary conditions | Periodic (X) Periodic (Y) Fixed (Z) [37] |
| Ensemble | *NVE* (implantation), *NPT* (annealing) |
| Dimensions | 9.6856 nm × 9.8743 nm × 18.814 nm |
| Timestep | Adaptive timestep (implantation),<br>0.5 fs (annealing) [38] |
| Thermostatic setup | Berendsen (293 K) [26] |
| Implantation region | 6 nm × 6 nm |
| Incident angle | 7° (away from the plane (001)) [27] |
| Ion implantation energy | 2 keV |
| Ion implantation fluence (Number of projectiles) | $1 \times 10^{14}$ $cm^{-2}$ (36), $2 \times 10^{14}$ $cm^{-2}$ (72), $3 \times 10^{14}$ $cm^{-2}$ (108), $4 \times 10^{14}$ $cm^{-2}$ (144), $5 \times 10^{14}$ $cm^{-2}$ (180) |
| Annealing temperatures | 1373K |

## 2.2 Defect identification

In the monoclinic crystal system of $\beta$-$Ga_2O_3$, there are two inequivalent Ga atoms: Ga1 occupying tetrahedral sites (coordination number = 4) and Ga2 residing in octahedral sites (coordination number = 6). Additionally, three distinct O atoms are present: O1, O2 (both with coordination = 3), and O3 (coordination = 4). A detailed view of the local structure projected along the (010) plane, as illustrated in Fig 2(a), reveals the atomic arrangement and bonding configurations at the five lattice sites. The Ga and O atoms are periodically arranged along the [010] direction, forming distinct atomic columns. The bonding between Ga and O creates channels extending parallel to the [010] direction, which are categorized as Channel 1 to Channel 4 according to their structural characteristics. In $\beta$-$Ga_2O_3$, interstitial atoms are typically situated within these channels, while vacancies tend to be located inside the atomic columns.

Currently, a common method for identifying point defects is the WS defect analysis. This approach involves partitioning the reference model into Voronoi cells. By comparing the defective model with the reference model, an occupancy of one atom per cell corresponds to a regular lattice atom (Occupancy = 1), multiple atoms in one cell indicate an interstitial defect (Occupancy > 1), and an empty cell signifies a vacancy (Occupancy = 0). However, due to the low-symmetry of the $\beta$-$Ga_2O_3$ crystal structure, when observed along the (010) plane (Fig. 2(b)), the Voronoi cells assigned to each atom vary significantly in volume, and the cell boundaries frequently lie within the channels. The interstitial atoms are likely located near these boundaries, resulting in misidentification by the WS method. To address this limitation, we have developed a Python-based algorithm specifically designed for identifying point defects in $\beta$-$Ga_2O_3$.

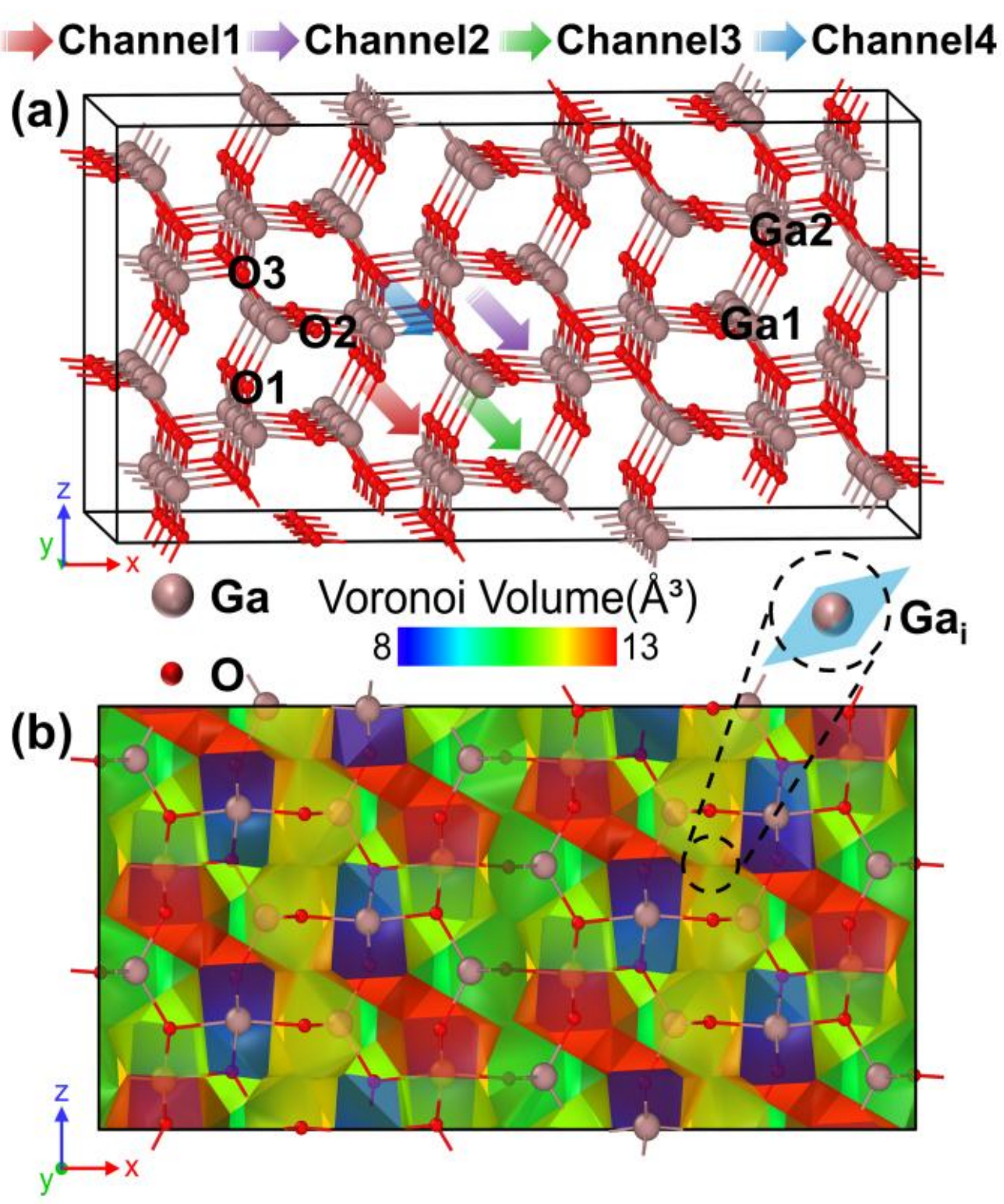

**Fig. 2.** Limitations of the WS Method in Identifying Interstitial Atoms in Non-centrosymmetric $\beta$-$Ga_2O_3$. (a) Atomic-scale structure of $\beta$-$Ga_2O_3$ viewed from the (010) plane, showing the five crystallographic sites (Ga1, Ga2, O1, O2, O3) and the channel structures indicated by arrows of four different colors. (b) Voronoi cells associated with each atom, colored by cell volume (8–13 Å$^3$), showing the prevalent location of cell boundaries within the structural channels. The dashed box indicates that the interstitial atoms are likely located at the Voronoi cell boundaries (indicated by blue planes).

Our point defect identification algorithm, implemented in Python, processes OVITO-readable data files containing atomic IDs, types, and *x*-, *y*- and *z*-coordinates. The *x*- and *z*-coordinates are utilized for DBSCAN clustering and similarity matching to distinguish between interstitial and lattice atoms, while the *y*-coordinate is employed to assess vacancy formation within clustered atomic groups. The algorithm outputs an OVITO-compatible data file, where visualization of interstitial and lattice atoms is achieved by modifying the atomic types in the input file. Vacancies are represented by creating new "atomic" entries with unique IDs, types, and coordinates. The workflow of the algorithm is shown in the supplementary material Figure S1.

Due to the arrangement of atoms in $\beta$-$Ga_2O_3$ forming atomic columns along the [010] direction, a dimensionality reduction approach was employed by projecting all atomic coordinates onto the *xz*-plane (corresponding to the (010) crystallographic plane). The DBSCAN algorithm was then applied to cluster closely neighboring atoms into the same group (i.e., belonging to the same atomic column), while filtering out interstitial atoms. Ga and O atoms were subjected to clustering analysis separately. To account for the influence of lattice thermal vibrations, the neighborhood radius (ε) was set as the maximum 2D Euclidean distance in the *xz*-plane between any two atoms within the same atomic column during the relaxation process ($\varepsilon_{Ga}$ = 0.494 Å, $\varepsilon_{O}$ = 0.386 Å; parameters are detailed in the supplementary material Figure S3). The minimum number of samples was set to three atoms (a configuration with very few atoms is likely an interstitial cluster, while cases with more interstitial atoms forming a column were further screened by subsequent similarity matching). Noise points identified by the algorithm were regarded as interstitial atoms. This method preliminarily achieved the distinction between lattice site atoms and interstitial atoms.

DBSCAN clustering was applied to both the relaxed perfect model (Fig. 3(a)) and the ion-implanted defective model (Fig. 3(b)). The cluster center for each atomic column was defined as the average position of the atomic coordinates within the *xz*-plane (represented by pentagrams for Ga centers and circles for O centers in Fig. 3(c) and (d), respectively; see the supplementary material Figure S3 for further details). Subsequent pattern recognition was performed based on these cluster centers in the *xz*-plane. Ga and O clusters that coordinate with each other are considered neighboring clusters (dashed lines in Fig. 3(c) and (d) indicate neighboring relationships). As shown in the enlarged views in Fig. 3(c) and (d), the perfect model exhibits a clear neighbor structure. Without interference from defects, clusters at five crystallographic sites can be accurately identified based on their neighbor configurations: Ga1 clusters have three neighbors, Ga2 clusters have four neighbors, and both O1 and O2 clusters have two neighbors. O1 clusters are distinguished by neighbor vectors tilted toward the *z*-axis, while O2 clusters are identified by vectors tilted toward the *x*-axis; O3 clusters with three neighbors are also present.

To achieve reliable identification under high defect density, it is necessary to screen for cases where multiple interstitial atoms are arranged in atomic columns based on the structural similarity between clusters. For the defective model, each cluster center obtained from clustering was compared with its counterpart in the perfect model. The comparison was based on the angular difference ($\Delta\theta$) in the distribution of neighbor clusters relative to the central cluster, along with the number of coordinating atoms ($N$). Based on these parameters, Formula 1 employs a sigmoid function to compute the similarity between clusters from the two models, which is subsequently used to match the cluster sites. This procedure established a unique one-to-one correspondence between the cluster sites of the two models:

$$S = \frac{1}{1 + \exp\left(-\frac{1}{M}\sum_{i=1}^{M}\left[\cos(\Delta\theta_i) \times \frac{N_{\text{defect},i}}{N_{\text{perfect},i}}\right]\right)}, \quad (1)$$

where, $S$ represents the structural similarity score between spatially proximate cluster centers in the defective model and the perfect model, $M$ denotes the number of neighboring clusters in the defect model, $\Delta\theta_i$ represents the angular difference between the $i$-th neighboring cluster in the defect model and its angularly closest counterpart in the perfect model, measured relative to the central cluster; $N_{defect,i}$ and $N_{perfect,i}$ correspond to the coordination numbers (Cutoff = 2.5 Å, the cutoff distance was chosen near the valley between the first and second peaks in the Ga-O RDF [28],) of the $i$-th neighboring cluster in the defect and perfect models, respectively. The similarity score increases as the angular distribution and coordination numbers of the neighboring clusters in the defect model more closely approximate those in the perfect model.

The results (Fig. 3(e) and (f)) show that the algorithm successfully distinguishes defect atoms from lattice-site atoms, identifies the crystal structure, and recognizes the five crystallographic sites. Finally, by tracking the immutable atomic IDs throughout the simulation and reassigning the atom types accordingly, we generated a data file for visualization in OVITO.

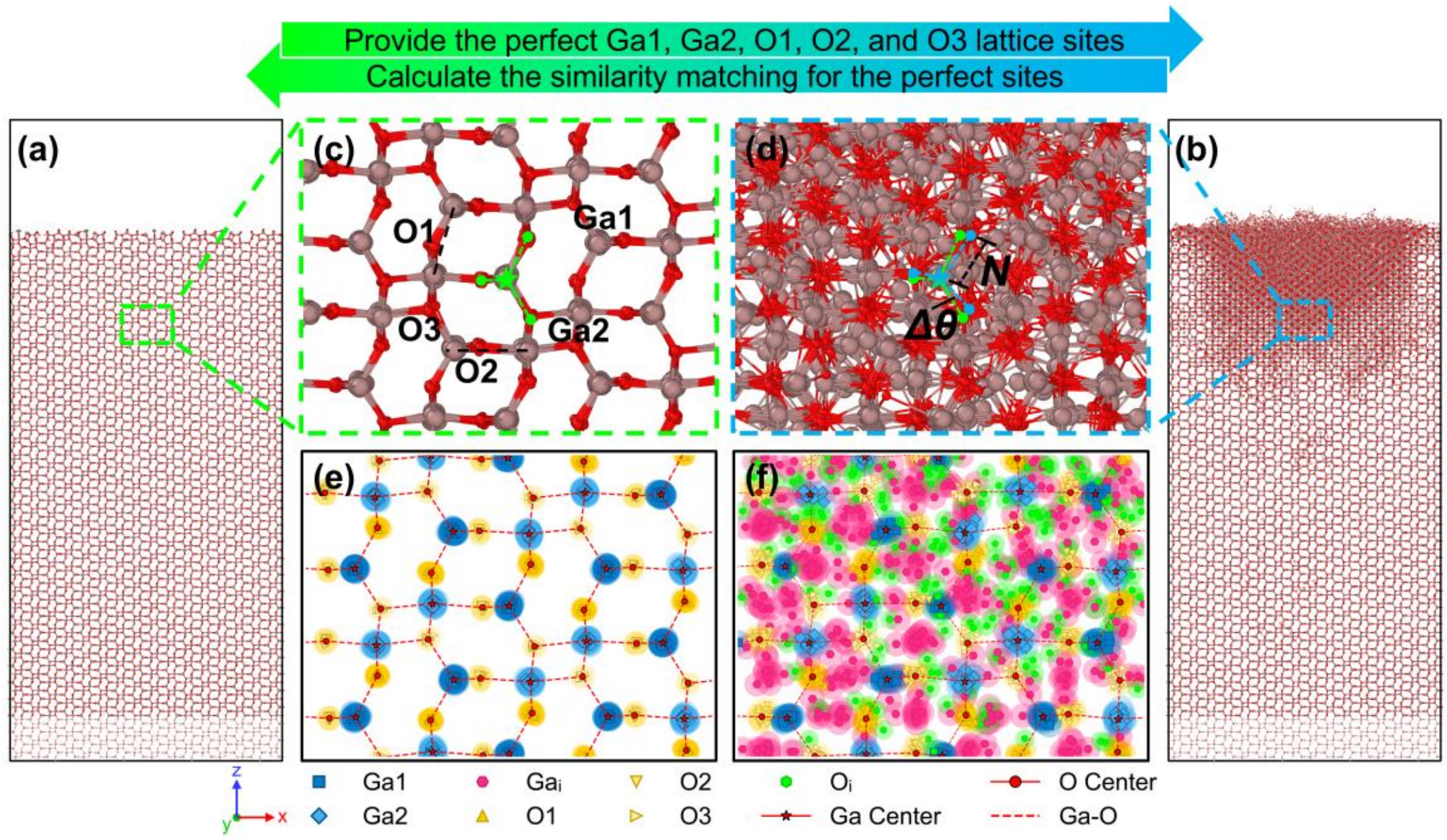

**Fig. 3.** (a) The relaxed perfect model at 293 K. (b) Defect model after ion implantation with a fluence of 3 × $10^{14}$ $cm^{-2}$. (c) Magnified view of the perfect model,

where green stars indicate Ga cluster centers, green circles denote O cluster centers (the centers represent the average coordinate positions of atoms clustered in the same group in the *xz*-plane. See the supplementary material Figure S3 for details), green dashed lines represent neighboring structures between clusters, and black dashed lines show the vectors formed by connecting O1/2 neighboring clusters. (d) Magnified view of the defect model, where blue-green gradient stars mark Ga cluster positions overlapping with the perfect model, blue circles indicate O cluster centers in the defect model, blue dashed lines depict neighboring structures between clusters, and compared to the perfect model, $\Delta\theta$ represents the angular difference in neighbor positions, and $N$ denotes the number of coordinating atoms between two clusters. (e,f) DBSCAN clustering results for the perfect and defect models, respectively. Symbols represent atom types and cluster centers, with outer circles indicating the clustering radius (ε).

To avoid misjudgments caused by lattice thermal vibrations, a comparison was performed at 0 K using single-ion implantation between the WS method and the DBSCAN method. The WS method (Fig. 4(a)) exhibited clear misclassifications, incorrectly assigning interstitial atoms as lattice-site atoms (occupancy = 1), the supplementary material Figure S11(a) also reveals the limitation of the WS method in identifying low-symmetry lattice, leading to an underestimation of lattice damage. In contrast, DBSCAN method (Fig. 4(b)) effectively corrected these errors, successfully identifying interstitial atoms and distinguishing atoms occupying different lattice sites (the method proves effective even in the presence of significant atomic thermal vibrations and a high defect density, with supporting test results provided in the supplementary material Figures S4 and S5). Vacancy identification was accomplished by analyzing the distances between adjacent atoms along the [010] direction within each atomic column; details are provided in the supplementary material Figure S3. Fig. 4(c) illustrates five distinct vacancy configurations ($V_{Ga1}$, $V_{Ga2}$, $V_{O1}$, $V_{O2}$, $V_{O3}$), as well as eight interstitial sites identified in $Ga_i$ (designated $Ga_{ia}$ to $Ga_{ih}$, following the nomenclature in Ref. [13]). Based on the approach described in Ref. [17] with appropriate modifications, the specific type of each interstitial site was determined by analyzing the coordination environment of each $Ga_i$ with nearby O atoms or vacancies (O1, O2, O3) and its relative position within the (010) plane. This was achieved using multiple cutoff radii ranging from 2.2 Å to 2.6 Å (a range sufficient to include all coordinating O atoms around $Ga_i$) with a step size of 0.05 Å; the final site type was defined as the interstitial type most frequently identified across this cutoff-radius range. Detailed atomic configurations and the identification methodology of the eight $Ga_i$ sites are provided in the supplementary material Figure S6. In particular, $Ga_i$ located at these interstitial sites often exists in the form of composite defect structures with adjacent $V_{Ga}$. When bonds are created using a cutoff radius of 2.5 Å, these $Ga_i$ spontaneously form bonds with nearby $V_{Ga}$ (composite structures are shown in the dashed boxes in the supplementary material Figure S6; all bonds connected to vacancies are not real chemical bonds and are used solely to represent the composite configurations).

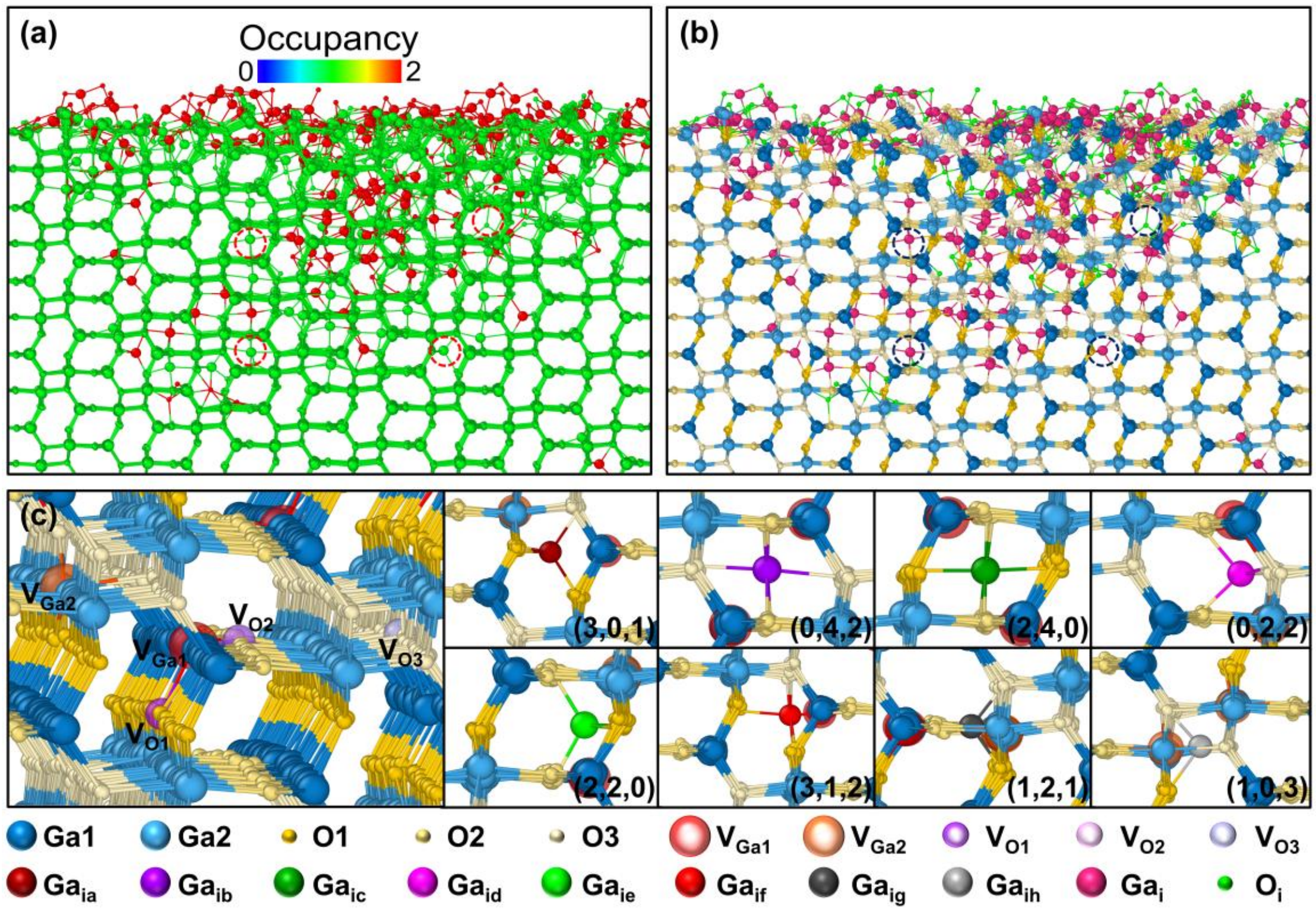

**Fig. 4.** (a) WS method identification: lattice atoms (green, Occupancy = 1) and interstitial atoms (red, Occupancy ≥ 2). Red circles highlight representative misidentified atoms. (b) DBSCAN method identification: Ga1/2 in dark and light blue; O1/2/3 in dark to light yellow; $Ga_i$ in pink; $O_i$ in green. Blue circles indicate representative corrected identifications. (c) Identifiable defect configurations: Five types of vacancy configurations ($V_{Ga1}$, $V_{Ga2}$, $V_{O1}$, $V_{O2}$, $V_{O3}$) identified by the spacing between adjacent atom pairs along the [010] direction within an atomic column (Methodology for vacancy detection is detailed in supplementary material Figure S3), and eight interstitial sites ($Ga_{ia}$ to $Ga_{ih}$) for $Ga_i$, which form composite configurations with surrounding vacancies (the bonds to vacancies are not real chemical bonds, but are used solely to represent the composite structure; detailed detection methods and structures of $Ga_i$ are provided in Figure S6), as determined by the coordination environment (values in parentheses indicate the numbers of O1, O2, O3 atoms, or vacancies).

## 3. Results and discussion

### 3.1 The effect of electronic stopping

During ion implantation, the energy loss resulting from collision cascades primarily consists of nuclear energy loss and electronic energy loss. Both the velocity and penetration depth of the implanted ions are influenced by electronic stopping. This section compares Stopping and Range of Ions in Matter (SRIM) [39] and MD simulation results to investigate the effect of electronic stopping during ion implantation, with the aim of improving the accuracy of subsequent simulations.

First, SRIM simulations were performed to study Ga and O ion implantation into $Ga_2O_3$ at various energies (the substrate was configured with a Ga-to-O atomic ratio of 2:3 and a mass density of 5.88 g/cm³ to approximate the properties of the $\beta$-phase.), as shown in Fig. 5(a). For heavy Ga ions, nuclear stopping is dominant. As energy increases, the rate of nuclear energy loss growth slows, while electronic energy loss increases more rapidly. For light O ions, nuclear stopping predominates at low energies, but electronic energy loss surpasses nuclear energy loss as energy increases. Both cases indicate that the influence of electronic stopping becomes more pronounced with increasing implantation energy. Therefore, the electronic stopping powers of O and Ga ions relative to Ga ions at different energies, as calculated by SRIM, were incorporated into the MD simulations.

Additionally, SRIM was used to simulate 1 million 2 keV Ga ions implanted into $Ga_2O_3$ (the SRIM simulation is based on the Monte Carlo method, and a larger ion count yields more accurate results. In this work, the ion beam was set at a tilt angle of 7°, and the displacement threshold energies for Ga and O were defined as 25 eV and 28 eV, respectively). The resulting ion concentration (per unit length) distribution was compared with that from an MD simulation using a fluence equivalent to $1 \times 10^{14}$ cm$^{-2}$ (corresponding to 36 Ga ions in the simulated system), as shown in Fig. 5(b). When electronic stopping is neglected, the implanted ion range distribution is more dispersed. In contrast, when electronic stopping is considered, the distribution becomes more concentrated and aligns more closely with the SRIM results, primarily distributed between 25–35 Å in depth. The trajectory line indicates that a Ga ion reaches a range of approximately 150 Å due to the channeling effect.

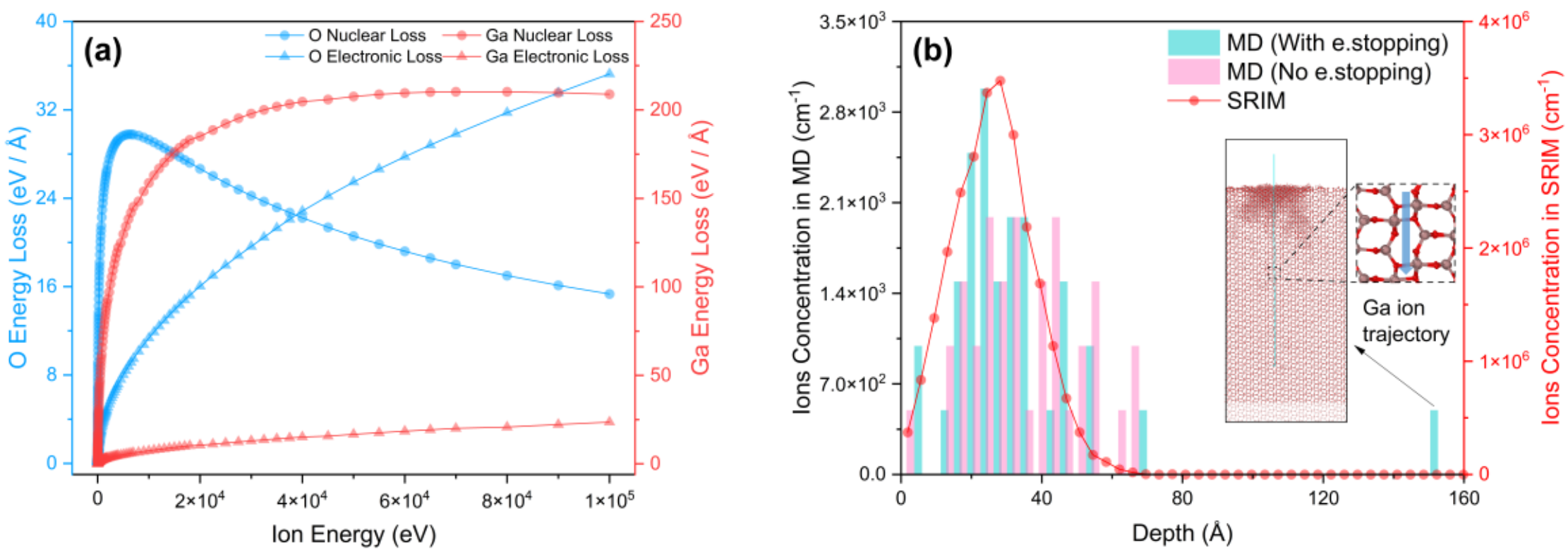

**Fig. 5.** (a) SRIM-simulated energy loss of Ga and O ions versus implantation energy. (b) Comparison of implanted ion concentration (per unit length) from SRIM and MD simulations, the inset shows the implanted ion trapped at a depth of approximately 150 Å due to the channeling effect, and the blue line represents the trajectory of the incident ion.

The influence of electronic stopping on ion ranges further affects the simulated temperature and defect generation. Therefore, we analyzed single-ion implantation at 0 K (to avoid the effects of lattice thermal vibrations) in our MD simulations (see the supplementary Videos 1 and 2 for dynamic trajectory comparisons). As shown in Fig. 6(a), upon interaction with the substrate, the implanted ion transfers a large amount of kinetic energy to the substrate, inducing a temperature rise in the Newtonian layer over an extremely short timescale (thermal-spike concept). The collision cascade stage occurs within approximately 2.5 ps. During this stage, the temperature in the Newtonian layer rises, while the concentrations of the four point defects ($Ga_i$, $O_i$, $V_{Ga}$ and $V_O$, detected using the DBSCAN method) increase rapidly initially, then decrease quickly due to the self-annealing effect induced by the localized high temperature (Fig. 6(b)). After 2.5 ps, as the kinetic energy of the implanted ion is largely dissipated, the system enters a cooling stage: the temperature decreases gradually, and the overall temperature of the Newtonian layer is higher when electronic stopping is neglected (see the local temperature contour maps in the supplementary material Figure S7). The changes in defect concentrations also become relatively slow. For all four types of point defects, the evolution curves obtained without considering electronic stopping are shifted upward compared to those obtained with electronic stopping included, indicating that neglecting electronic stopping leads to an overestimation of defect production. Given the cumulative nature of ion implantation, this overestimation would be amplified with increasing fluence, thus introducing significant errors in the evaluation of lattice damage levels in $\beta$-$Ga_2O_3$ and the threshold of $\gamma$-phase transformation induced by ion implantation. Therefore, it is reasonable and necessary to account for electronic stopping in simulations of ion implantation processes.

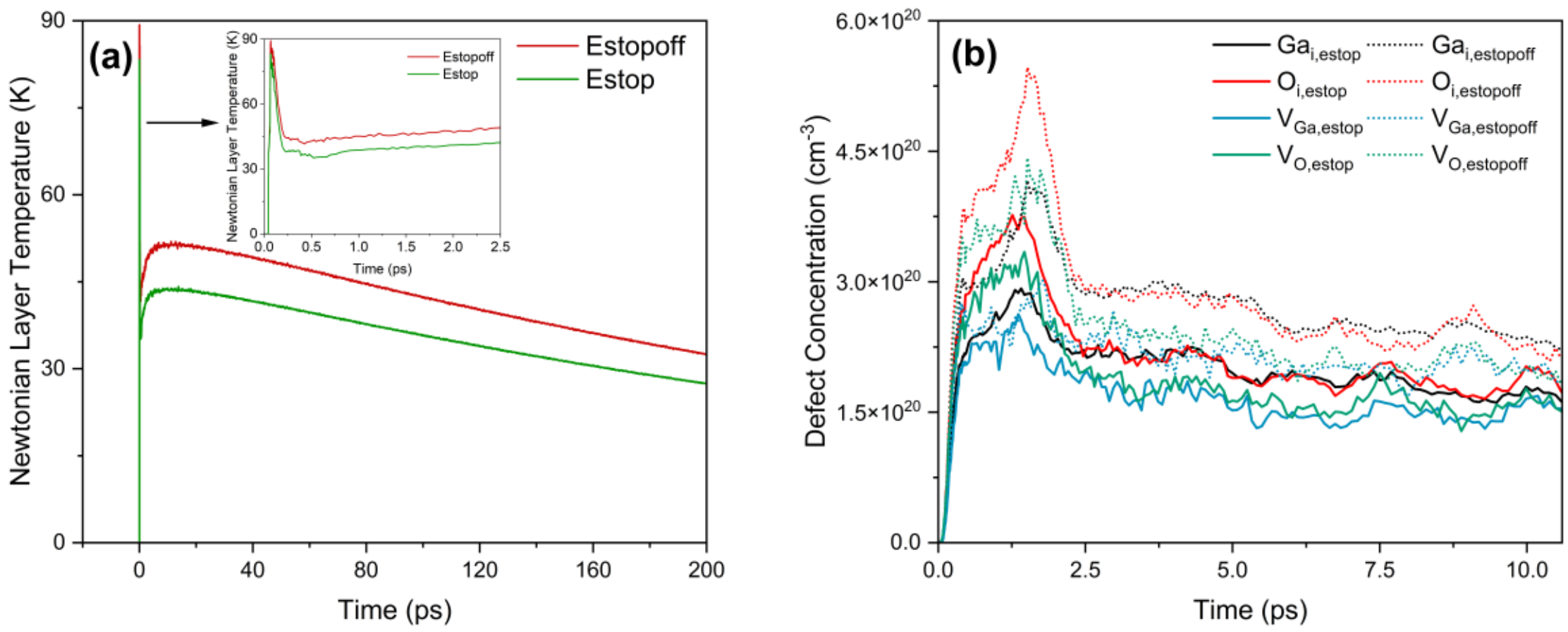

**Fig. 6.** Effect of electronic stopping in MD simulations of single-ion implantation at 0 K. (a) Newtonian layer temperature evolution, and the inset presents the temperature evolution within 2.5 ps. (b) Point defects evolution.

3.2 Ion Implantation and Annealing

This section focuses on the evolution mechanisms of various defects in $\beta$-$Ga_2O_3$ under ion implantation and subsequent annealing. Interstitial atoms (including all atoms displaced from the $\beta$-phase lattice), as the primary defects resulting from lattice damage produced in implantation-induced collision cascades, are investigated first. To avoid errors induced by surface damage from high-fluence ion implantation, the analysis of point defect evolution was confined to the crystal interior. The average point defect concentration was calculated for layers starting at depths of 0, 2, 4, 6, 8, and 10 Å and extending into the crystal interior. Fig. 7(a) shows the variation in the concentrations of different types of interstitials as a function of implantation fluence. The concentrations of $Ga_i$ and $O_i$ were recorded after the relaxation of each implantation step. Both $Ga_i$ and $O_i$ exhibit an increasing trend with fluence, though the rate of increase slows at higher fluences, suggesting the approach toward a saturation level. Notably, while single-ion implantation at 0 K resulted in a lower concentration of Ga-related defects than O-related defects (Fig. 6(b)), the opposite trend is observed at 293 K. Moreover, the concentration curve of $O_i$ shows greater fluctuation compared to that of $Ga_i$. This behavior may be attributed to the higher mobility of $O_i$ at elevated temperatures, facilitating their migration back to lattice sites. Subsequently, the distribution of $Ga_i$ among the eight interstitial sites (from $Ga_{ia}$ to $Ga_{ih}$ ,which may be associated with the phase transition, will be further discussed in subsequent sections) is analyzed, as shown in Fig. 7(b). $Ga_{if}$ is the most frequently occupied interstitial site during implantation. The concentrations of $Ga_{ia}$, $Ga_{ib}$, and $Ga_{ic}$ are comparable, ranking second only to $Ga_{if}$, while $Ga_{ie}$, $Ga_{ig}$, and $Ga_{ih}$ are the least occupied sites. The evolution of vacancies can be found in the supplementary material Figure S8.

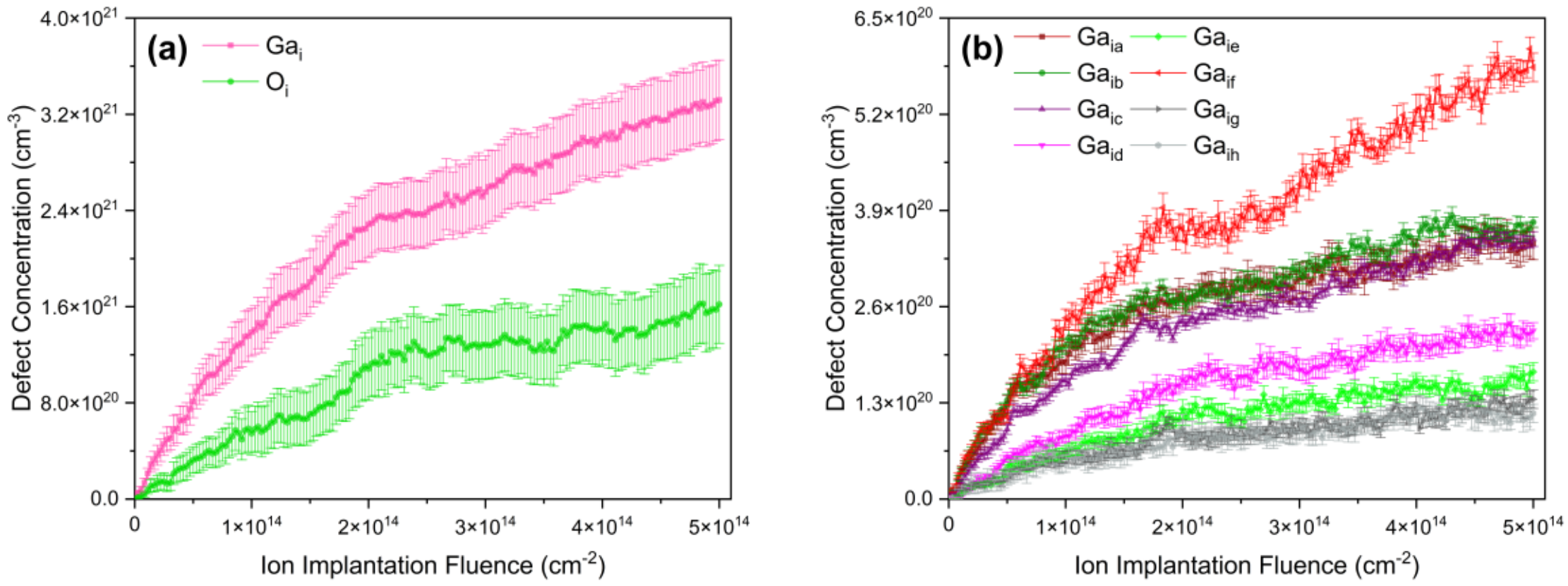

**Fig. 7.** (a) Evolution of $Ga_i$ and $O_i$ as a function of implantation fluence. (b) Evolution of $Ga_{ia}$ to $Ga_{ih}$ (sub-sites of $Ga_i$) as a function of implantation fluence. Error bars show standard deviations.

Annealing effectively repairs lattice damage induced by ion implantation. Based on an implantation fluence of $2 \times 10^{14}$ $cm^{-2}$, annealing simulations were performed at three different temperatures: 1273 K, 1373 K, and 1473 K. The optimal annealing temperature was determined to be 1373 K, which is consistent with the experimental annealing temperatures reported in Refs. [7,8]. The results are summarized in Fig. 8. Significant defect recovery was observed after annealing at both 1273 K and 1373 K, as evidenced by a pronounced reduction in the concentration of interstitial atoms. The $Ga_i$ concentration reached its minimum after annealing at 1373 K. Furthermore, the $O_i$ concentration decreased remarkably at these two temperatures, further confirming that O-related defects exhibit higher sensitivity to temperature and can more readily migrate back to lattice sites at elevated temperatures, thereby promoting the recrystallization of the O-sublattice. In contrast, annealing at 1473 K introduced a substantial number of surface defects, leading to surface amorphization, which failed to achieve the annealing objectives. The concentration profile of vacancies during the annealing tests can be found in the supplementary material Figure S9.

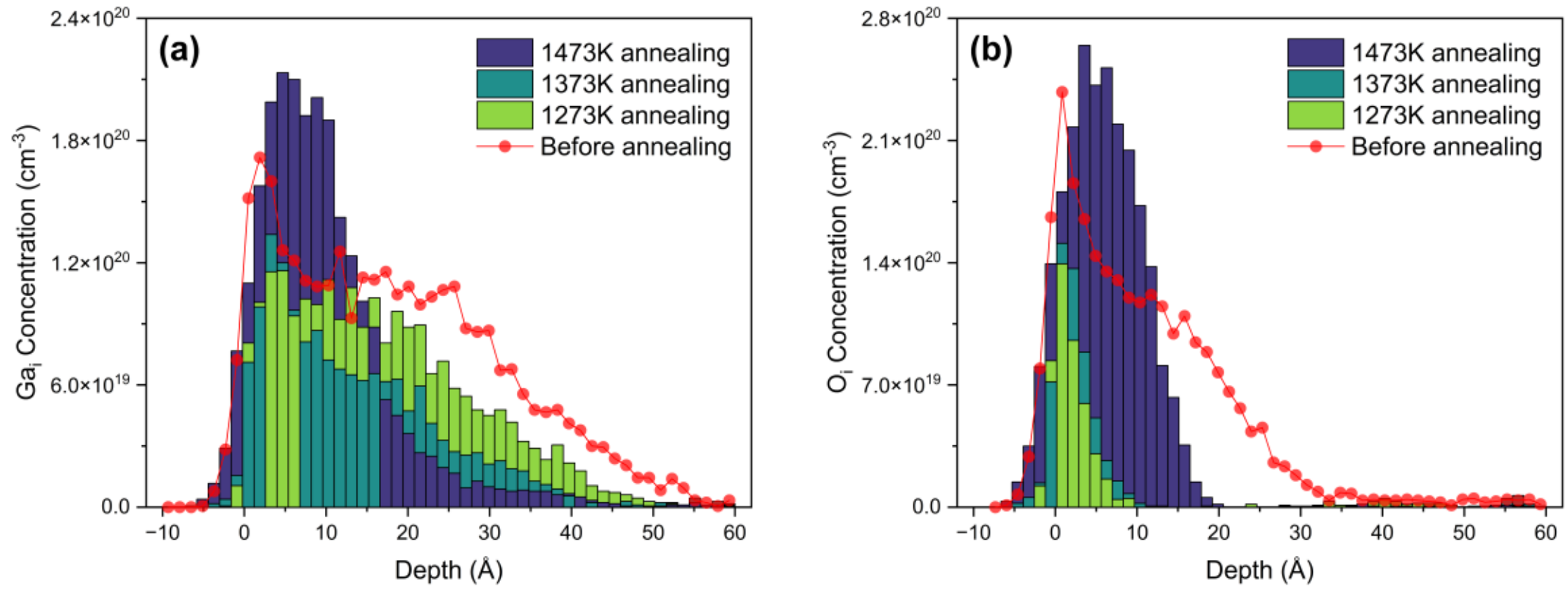

**Fig. 8.** Annealing temperature test at $2 \times 10^{14}$ cm$^{-2}$. (a) $Ga_i$ concentration profile before and after annealing. (b) $O_i$ concentration profile before and after annealing.

To further analyze the defect evolution during ion implantation and annealing, Fig. 9 presents the concentration profiles of $Ga_i$ and $O_i$ after ion implantation at five different fluences followed by annealing at 1373 K (the vacancy concentration profile corresponding to this stage can be found in the supplementary material Figure S10). The results indicate that high-fluence ion implantation induces lattice damage near the surface, as evidenced by the accumulation of interstitial atoms in the near-surface region. Compared with $Ga_i$, $O_i$ exhibits a stronger tendency to accumulate at the surface. However, the O-sublattice shows superior recovery after annealing. In contrast, a significant concentration of $Ga_i$ remains after annealing. Moreover, the three-dimensional distribution of interstitials (the supplementary material Figure S13) reveals that $Ga_i$ become orderly arranged after annealing, suggesting that most of the residual $Ga_i$ migrate to relatively stable interstitial sites (within the range $Ga_{ia}$ to $Ga_{ih}$) during annealing, rather than remaining in disordered configurations. Based on the concentration profiles, the depth region of 10–50 Å was defined as the high-defect zone within the crystal, where PRDF analysis was performed. Due to lattice damage induced by implantation, the atomic arrangement inside the crystal becomes more disordered with increasing fluence, manifested as a decrease in PRDF peak intensity. Annealing partially repairs the damage. Specifically, the O-O PRDF indicates pronounced recovery of the O-sublattice, suggesting that the O-sublattice possesses relatively high rigidity [40].

In contrast, the Ga-Ga PRDF reveals a phase transformation from the $\beta$-phase to the $\gamma$-phase during implantation and annealing (the $\beta$-phase Ga-Ga PRDF shows a characteristic peak near 4.6 Å, while the $\gamma$-phase exhibits a monotonic increase between 4.6 and 5.2 Å [4]). With increasing implantation fluence, the $\gamma$-phase features become more pronounced. Although annealing promotes partial recovery of the $\beta$-phase, the Ga-Ga PRDF curves still retain enhanced $\gamma$-phase characteristics after high-fluence annealing, indicating that the Ga-sublattice is more prone to phase transformation during ion implantation and subsequent annealing. In the supplementary material Appendix F, the lattice damage and the phase transition threshold are discussed in detail using displacements per atom (DPA) and the Pearson correlation coefficient (Pr) [4]. The Ga-sublattice undergoes $\beta$-phase lattice damage and $\gamma$-phase formation during ion implantation, while the annealing process facilitates $\beta$-phase recovery while also promoting $\gamma$-phase formation. Consequently, an irreversible phase transition threshold corresponding to the $\beta \rightarrow \gamma$ transformation in the Ga-sublattice is established.

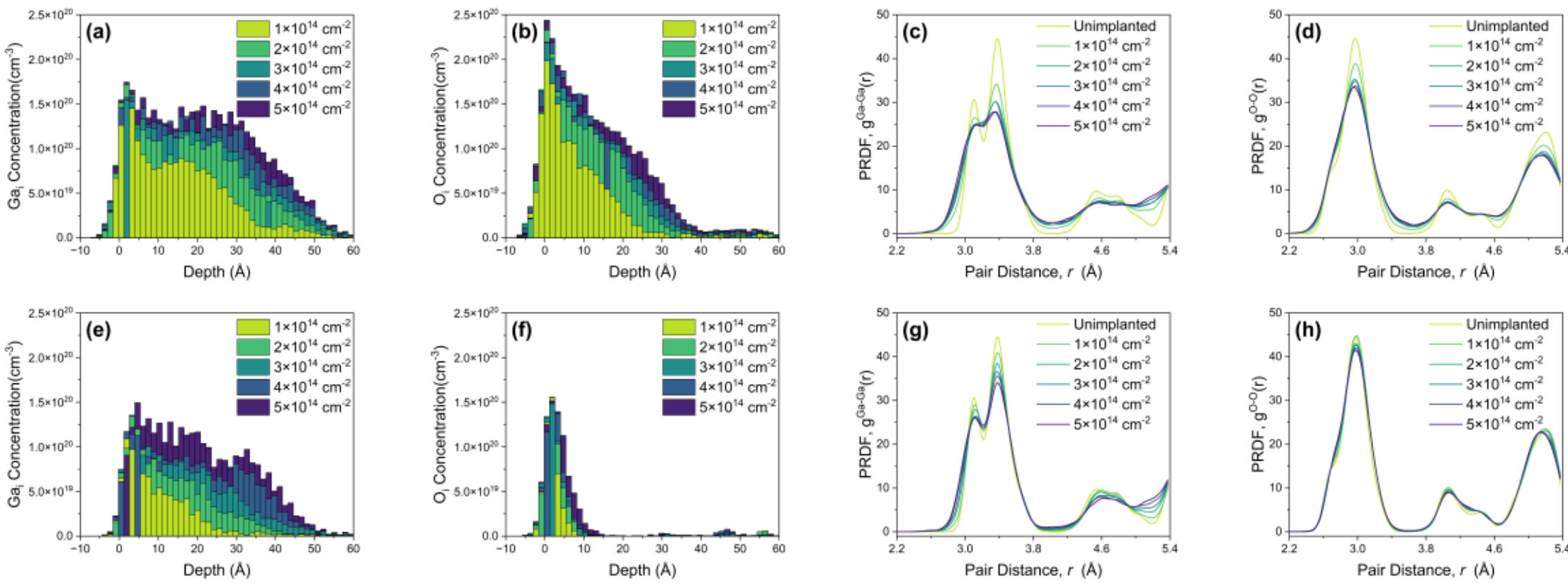

**Fig. 9.** (a–d) As-implanted state and (e–h) after annealing at 1373 K. (a, e) $Ga_i$ concentration profile. (b, f) $O_i$ concentration profile. (c, g) Ga-Ga PRDF profiles. (d, h) O-O PRDF profiles.

Subsequently, hydrostatic stress [41] and the distribution of $Ga_i$ at the $Ga_{ia}$ to $Ga_{ih}$ sites were characterized within regions with depths less than 50 Å along the (010) plane after implantation and annealing at five different fluences, as shown in Fig. 10 (first and third rows). The stress contour maps reveal that the implanted system is predominantly under tensile stress, and this tensile stress distribution becomes more pronounced after annealing. This can be attributed to the fact that after implantation, a fraction of $Ga_i$ atoms occupy the $Ga_{ia}$ to $Ga_{ih}$ sites. These $Ga_i$, together with neighboring $V_{Ga}$, form composite defect configurations, thereby inducing tensile stress (stress $> 0$). In contrast, $O_i$ are distributed in a disordered state, causing lattice distortion and resulting in compressive stress (stress $< 0$), as detailed in the atomic stress analysis provided in the supplementary material Figure S14. After annealing, most $O_i$ migrate back to the O-sublattice, while $Ga_i$ predominantly relocate to the $Ga_{ia}$ to $Ga_{ih}$ sites, as illustrated in Fig. 10 (second and fourth rows). The spatial distribution of these $Ga_i$ closely correlates with the regions of tensile stress, which explains the enhanced tensile stress after annealing (in the supplementary material Figure S16 presents a detailed analysis of the system pressure and $Ga_i$ stress following annealing at a fluence of $5 \times 10^{14}$ cm$^{-2}$). These findings highlight the irradiation resistance of $\beta$-$Ga_2O_3$: although $O_i$ tends to exist in a disordered state after implantation, annealing effectively promotes its recovery. In comparison, $Ga_i$ preferentially migrate to the $Ga_{ia}$ to $Ga_{ih}$ sites, which are associated with the formation of the $\gamma$-phase.

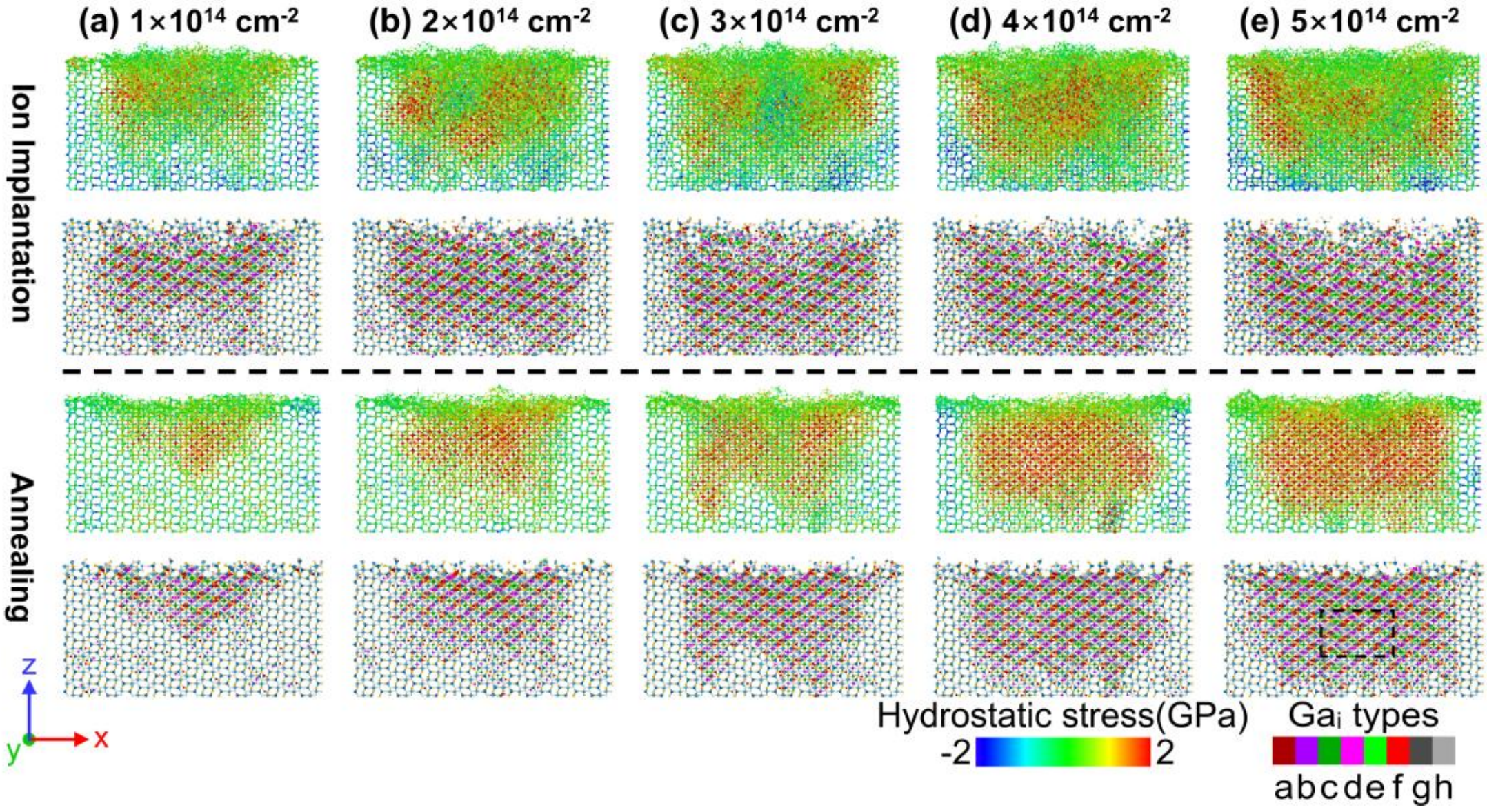

**Fig. 10.** Hydrostatic stress contour maps (rows 1 & 3) and corresponding $Ga_i$ defect morphology (rows 2 & 4) under five different implantation fluences. The dashed line separates the as-implanted state (top) from the state after annealing at 1373 K (bottom). The hydrostatic stress color bar ranges from -2 to 2 GPa. In the morphology plots, only $Ga_i$ at interstitial sites ($Ga_{ia}$ to $Ga_{ih}$) are shown and colored according to their type, alongside lattice atoms for reference.

To further investigate the defect structures in $\beta$-$Ga_2O_3$, a cross-sectional slice 20 Å thick was extracted from the central region of the (010) plane of the model annealed at a fluence of $5 \times 10^{14}$ cm$^{-2}$. For clarity, only $Ga_i$ and $V_{Ga}$ are displayed, as shown in Fig. 11(a). The results demonstrate that Ga-related defects primarily exist in the form of $Ga_i$-$V_{Ga}$ complexes. Among these, the most frequently observed configuration in $\beta$-$Ga_2O_3$ is a split-vacancy chain arranged as $V_{Ga1}$-$Ga_{ib}$-$V_{Ga1}$-$Ga_{ic}$-$V_{Ga1}$ ($V_{Ga}^{ibc}$)[13] which is marked by a red dashed box in Fig. 11(a). To explore the correlation between the $Ga_{ia}$ to $Ga_{ih}$ sites and the phase transformation, Fig. 11(b) presents a magnified view of the area within the black dashed box in Fig. 10, revealing numerous $\gamma$-phase regions characterized by the (110) planes of the spinel structure [8,42,43]. After annealing, as shown in the four representative spinel structures selected in Figs. 11(c–e), the phase transformation is driven by defect accumulation through four distinct modes. Furthermore, The supplementary material Figure S14 indicates that the proportions of both $V_{Ga}$ and the $Ga_i$ sites (from $Ga_{ia}$ to $Ga_{ih}$) after annealing gradually increase with rising implantation fluence. The complexing and accumulation of $V_{Ga}$ and $Ga_i$ further promote the initiation of the phase transformation.

Since $Ga_i$ occupy different tetrahedral and octahedral interstitial sites (the supplementary material Figure S6), they are classified based on their coordination polyhedra: $Ga_{ia}$, $Ga_{id}$, $Ga_{ie}$, and $Ga_{ig}$ belong to the $Ga_i1$ type (occupying tetrahedral interstitial sites), while $Ga_{ib}$, $Ga_{ic}$, and $Ga_{if}$ are categorized as $Ga_i2$ (occupying octahedral interstitial sites). By tracking atomic IDs, the migration behavior of $Ga_i$ after annealing at different fluences was analyzed. The results show that most $Ga_i$ migrate to lattice sites (Ga1 and Ga2 are associated with the recovery of the $\beta$-phase) or interstitial sites ($Ga_i1$ and $Ga_i2$ are associated with the formation of the $\gamma$-phase.) after annealing, with only a small fraction remaining in their original metastable $Ga_i$ state (amorphous interstitial sites resulting from lattice damage). Compared to tetrahedral lattice and interstitial sites (Ga1, $Ga_i1$), $Ga_i$ exhibits a stronger tendency to migrate to octahedral sites (Ga2, $Ga_i2$). As the implantation fluence increases, the proportion of $Ga_i$ migrating to interstitial sites rises, while the proportion occupying lattice sites decreases. These findings provide further insight into the atomic-scale mechanism of the $\beta$-to-$\gamma$ phase transformation in $\beta$-$Ga_2O_3$ after annealing and suggest that the irreversibility of the transformation is progressively enhanced with increasing fluence. Finally, the substitution rate of the implanted Ga ions after annealing at each fluence was quantified, as shown in Fig. 11(h), the migration behavior of $Ga_i$ is in high agreement with the phase transition mechanism illustrated in the supplementary material Figure S12(b), both demonstrating an annealing-irreversible phase transition threshold at a fluence of $4 \times 10^{14}$ $cm^{-2}$: with increasing fluence, the occupancy of interstitial sites increases, the occupancy of lattice sites declines, and the fraction remaining in the $Ga_i$ state becomes negligible. This phenomenon further confirms that two trends exist during the annealing of $\beta$-$Ga_2O_3$: recovery of the $\beta$-phase and transformation to the $\gamma$-phase. At low fluence, annealing is dominated by the recovery of the $\beta$-phase, whereas with increasing fluence, the $\gamma$-phase transformation becomes predominant, thereby leading to the irreversibility of the phase transition. The migration behavior of $Ga_i$ provides a theoretical basis for the residual $\gamma$-phase observed in experiments [8] after high-fluence implantation and annealing.

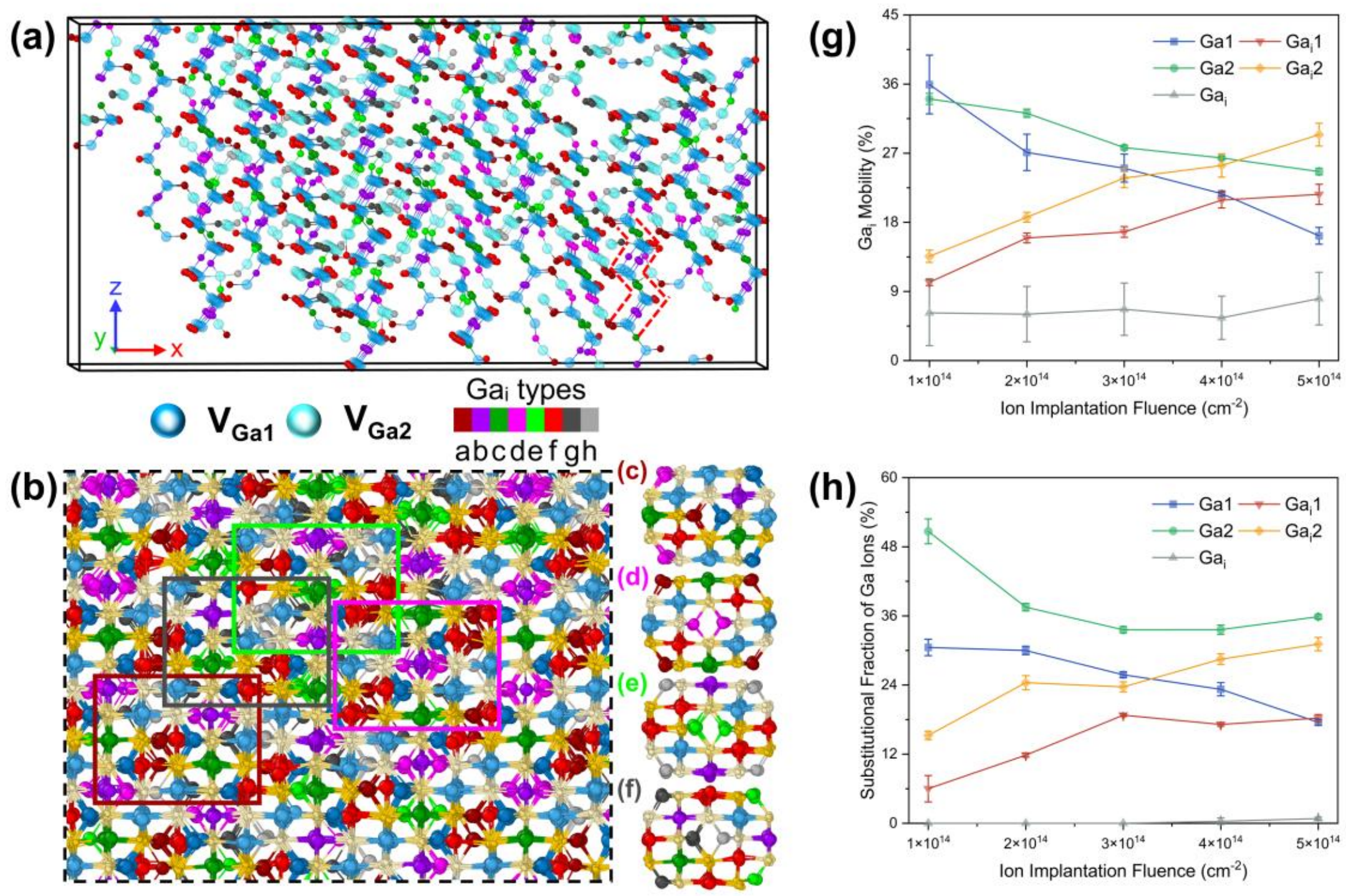

**Fig. 11.** (a) Cross-sectional view (20 Å thick) of the sample annealed at a fluence of $5 \times 10^{14}$ cm$^{-2}$, showing $Ga_i$-$V_{Ga}$ complex structures. The red dashed line indicates a representative defect complex. (b) Magnified view of the annealed model at $5 \times 10^{14}$ cm$^{-2}$, where solid boxes in different colors highlight the (110) planes of the spinel structure formed by different $Ga_i$ configurations (c–f). (g) Migration behavior of $Ga_i$ after annealing at five different implantation fluences. (h) Substitution rate of implanted Ga ions after annealing at each of the five fluences. Error bars show standard deviations.

## 4.Conclusion

In this study, a similarity-matching algorithm based on DBSCAN clustering was developed. This method effectively distinguishes point defects (interstitials and vacancies) from lattice atoms, enabling accurate identification even under thermal perturbations and at high defect densities.

Subsequently, by comparing SRIM simulations with MD simulations and performing single-ion implantation in MD, the influence of electronic stopping on the ion range was investigated. It was found that neglecting electronic stopping significantly overestimates the ion range, system temperature, and the concentrations of various defects.

Using MD simulations, a comprehensive atomic-scale analysis of the defect evolution mechanism during Ga ion implantation and subsequent annealing in $\beta$-$Ga_2O_3$ was performed. The results reveal that $Ga_i$ tends to occupy eight interstitial sites (from $Ga_{ia}$ to $Ga_{ih}$) during implantation and annealing, forming complex structures with adjacent $V_{Ga}$ that introduce tensile stress. These interstitial configurations are identified as the primary factors inducing the $\beta$-to-$\gamma$ phase transition in $\beta$-$Ga_2O_3$. Furthermore, the migration behavior of $Ga_i$ that with increasing fluences, the tendency for the $\gamma$-phase transformation in post-annealed $\beta$-$Ga_2O_3$ rises continuously, accompanied by a corresponding decrease in the trend of $\beta$-phase recovery, ultimately leading to an irreversible phase transition. In contrast, $O_i$ exhibits temperature-dependent migration behavior, and the O-sublattice demonstrates relatively high rigidity.

**CRediT authorship contribution statement**

**Huawen Li:** Conceptualization, Data curation, Formal analysis, Investigation, Software, Methodology, Visualization, Writing–original draft. **Mengzhi Yan:** Conceptualization, Methodology, Software, Resources, Validation, Writing - review & editing. **Zongwei Xu:** Funding acquisition, Project administration, Resources, Supervision. **Junlei Zhao:** Conceptualization, Supervision, Writing - review & editing. **Jiale Wang:** Investigation, Validation.

**Declaration of competing interests**

The authors declare that they have no known competing financial interests or personal relationships that could have appeared to influence the work reported in this paper.

**Acknowledgments**

This work was supported by the National Key Research and Development Program Project (2024YFF0726104).

**Supplementary materials**

Supplementary material related to this article can be found online at xxx.

# Supplementary Material: Machine-learning-guided molecular dynamics simulations of point defect evolution in $\beta$-$Ga_2O_3$ during ion implantation and annealing

Huawen Li[1], Mengzhi Yan[2], Zongwei Xu[1,*], Junlei Zhao[3], Jiale Wang[1]

[1]*State Key Laboratory of Precision Measuring Technology and Instruments, Laboratory of Micro/Nano Manufacturing Technology (MNMT), Tianjin University, Tianjin 300072, China*

[2]*Centre of Micro/Nano Manufacturing Technology (MNMT-Dublin), School of Mechanical and Materials Engineering, University College Dublin, Belfield, Dublin 4, D04 V1W8, Ireland*

[3]*The Hong Kong Microelectronics Research and Development Institute, Hong Kong, 999077, China*

[*]Corresponding Author: zongweixu@tju.edu.cn

**CONTENTS**

**Appendix A: Introduction to defect identification parameters**

The workflow of our point defect identification algorithm is illustrated in Figure S1. The input files can be data files exported from OVITO, which should include five columns of data: atom ID, atom type, and *x*-, *y*-, *z*- coordinates. Atom types must be specified as Ga and O, corresponding to type 1 and type 2, respectively. The Python script requires two inputs: a perfect reference model file (i.e., a relaxed structure only) and one or more defective model files after processing (e.g., ion implantation, laser modification, nano-machining, etc.). Multiple defective files can be provided with filenames ordered by an asterisk * followed by natural numbers in increasing order; the program will process them sequentially from the smallest to the largest number.

The program first applies periodic boundary handling to both the perfect and defect models, replicating atoms within 2.5 Å of the boundary to the opposite side of the simulation box. These replicated atoms serve only as auxiliary particles to complete the neighbor environment near the boundary and are not included in the final output. After DBSCAN clustering, atoms in the perfect model that are crystallographically close are grouped together. Neighbor analysis is then performed to determine the atom types within each group, generating a model containing Ga1/2 and O1/2/3 sites, which provides a reference lattice for the defect model. For the defect model, DBSCAN clustering is applied, and initially identified noise points are used to preliminarily screen irregularly arranged interstitial atoms. During neighbor analysis, these preliminarily identified interstitials are filtered out to obtain a relatively clean atomic environment for similarity matching with the perfect model. After matching, lattice atoms and interstitial atoms are further distinguished. Subsequently, vacancies are created within the retained lattice atom groups. The program outputs data files compatible with OVITO visualization, containing different types of vacancies, interstitials, and lattice atoms, distinguished by atom types in the data file.The identification of eight interstitial sites from $Ga_{ia}$ to $Ga_{ih}$ is performed by another Python script that further processes the output.*.data files. *Complete documentation and resources of the identification algorithm are open-source on GitHub at: https://github.com/L-H-W/A-DBSCAN-and-Similarity-Based-Algorithm-for--Ga2O3-Point-Defect-Identification.git.*

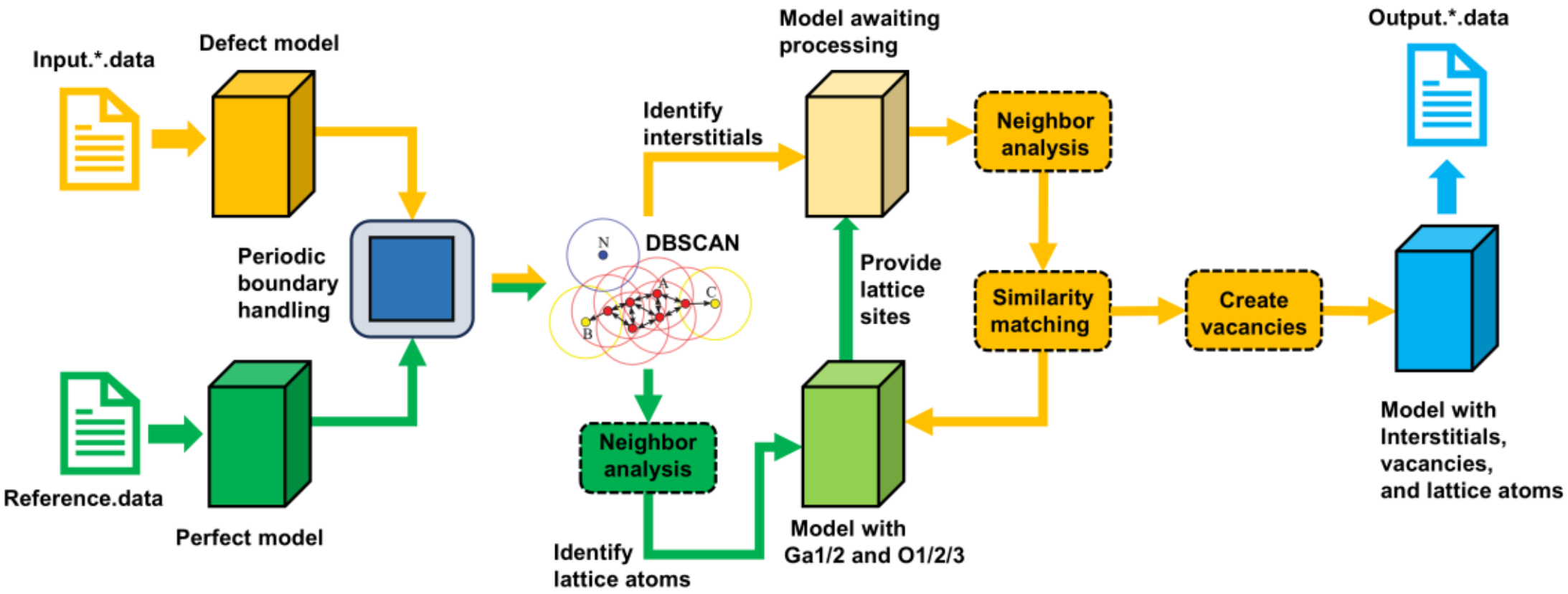

**Figure S1.** Workflow of the point defect identification algorithm, with color-coded arrows indicating the processed models.

$\beta$-$Ga_2O_3$ exhibits a monoclinic crystal structure belonging to the space group C2/m. As shown in Figure S2(a), the primitive unit cell features an angle of 103.8° between the [001] and [100] directions [1]. By rotating the primitive cell clockwise with the [010] axis as the rotation axis, the [001] direction is aligned with the *z*-axis, yielding an orthogonalized unit cell (Figure S2(b)). In the orthogonalized configuration, the *z*-axis corresponds to the [001] direction, the *y*-axis aligns with [010], and the *x*-axis deviates from [100] by 13.8° (a non-conventional orientation). Figures S2(c–e) present different views of the orthogonalized cell. The left view shows the projection onto the *yz*-plane, where Ga and O atoms at different crystallographic sites are projected to closely overlapping positions, complicating discrimination. In the top view (*yz*-plane), the projected coordinates of O1 and O3 atoms are nearly coincident. Consequently, performing atom clustering and type-grouping on these two planes is highly inefficient and challenging. In contrast, the front view (*xz*-plane) reveals a well-ordered periodic arrangement of atoms along the [010] direction, with a uniform inter-atomic spacing of 3.08572 Å between adjacent Ga/O atoms along [010], which facilitates the analysis of vacancy formation. Moreover, atoms belonging to the same crystallographic site exhibit closely clustered projections in the *xz*-plane, while those from different sites are well separated. This structural feature greatly favors effective clustering and atomic grouping. Therefore, our point-defect identification algorithm is developed based on the (010)-plane projection.

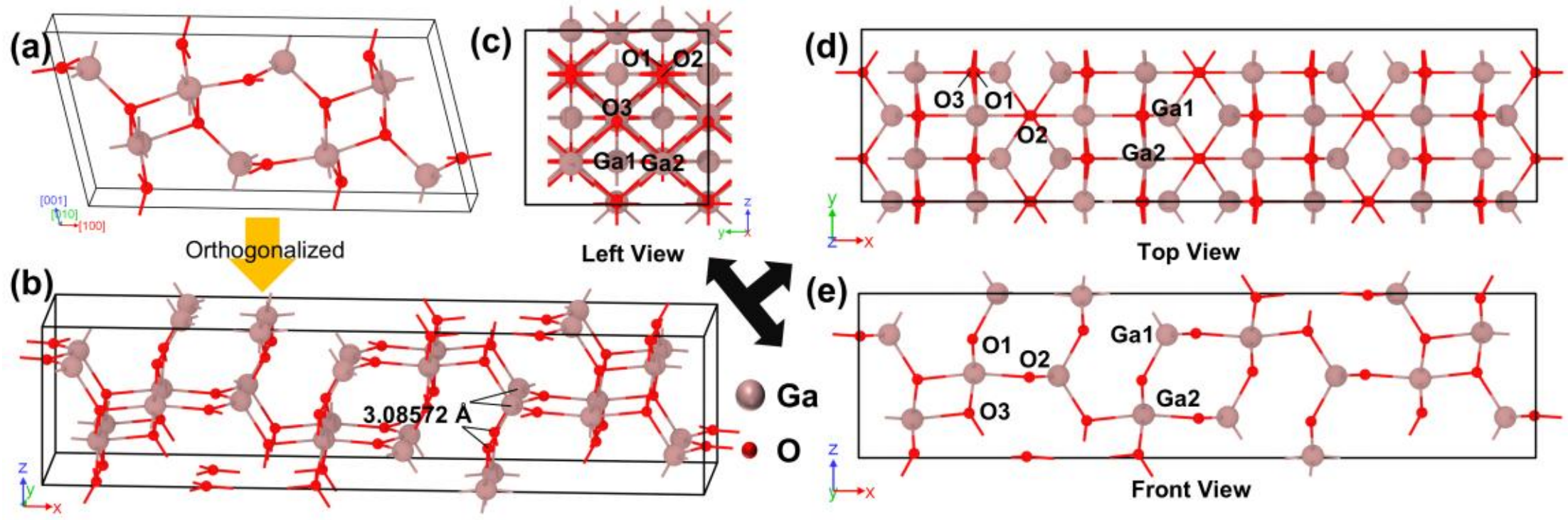

**Figure S2.** (a) Primitive unit cell of *β*-$Ga_2O_3$, with axes indicating its crystallographic directions. (b) Orthogonalized unit cell, the labeled distance corresponds to the spacing between adjacent Ga/O atoms arranged along the [010] direction. (d–e) Different viewing directions of the orthogonalized unit cell, in which atoms are labeled according to their crystallographic sites for Ga and O.

As shown in Figure S3, lattice thermal vibrations are present in *β*-$Ga_2O_3$ during MD simulations, leading to fluctuations in atomic positions after relaxation. However, the motion of atoms within the crystal does not deviate from the atomic columns along the [010] direction, and the primary fluctuations occur within the (010) crystallographic plane (i.e., the *xz*-plane). Therefore, to accurately cluster lattice atoms belonging to the same atomic column into one group, we adopted an approach where the maximum two-dimensional Euclidean distance between any two atoms within all atomic columns during the relaxation process is set as the clustering radius (the selection for Ga atoms is illustrated in Supplementary Figure S3(d), and similarly for O atoms). The currently obtained parameters ($\varepsilon_{Ga}$ = 0.494 Å, $\varepsilon_{O}$ = 0.386 Å) enable effective atomic identification at 293 K. The center of each cluster (represented by the red pentagram for the Ga center in Supplementary Figure S3(d), and similarly for the O center) is determined by averaging the coordinates of all atoms within the atomic column in the *xz*-plane. This center point is subsequently used for similarity matching judgments.

Regarding the vacancy formation mechanism illustrated in Figures S3(e) and (f), due to the relatively minor variation in the interatomic distance (dy) along the [010] direction on the *yz*-plane after relaxation, when vacancies are generated in an atomic column owing to the loss of atoms, the distance between the two remaining adjacent atoms in that column will approximate $n \times dy$ if $n-1$ atoms are missing ($n \geq 2$). Therefore, after filtering out interstitial atoms, vacancy identification is performed for each cluster (corresponding to retained lattice atomic columns) obtained via DBSCAN classification. The average distance between adjacent atoms along the [010] direction across all atomic columns over the entire relaxation process (3.085 Å) is adopted as the criterion for vacancy generation. Atoms within the atomic column are sorted according to their $y$-coordinates. If the difference in $y$-coordinates between two neighboring atoms approaches n × dy, n−1 vacancies are uniformly distributed between the two atoms by linear interpolation, with the vacancy type determined by the atomic species within the cluster. It should be noted that the clustering radius and vacancy generation parameters are not fixed and can be adjusted according to different relaxation conditions and specific requirements.

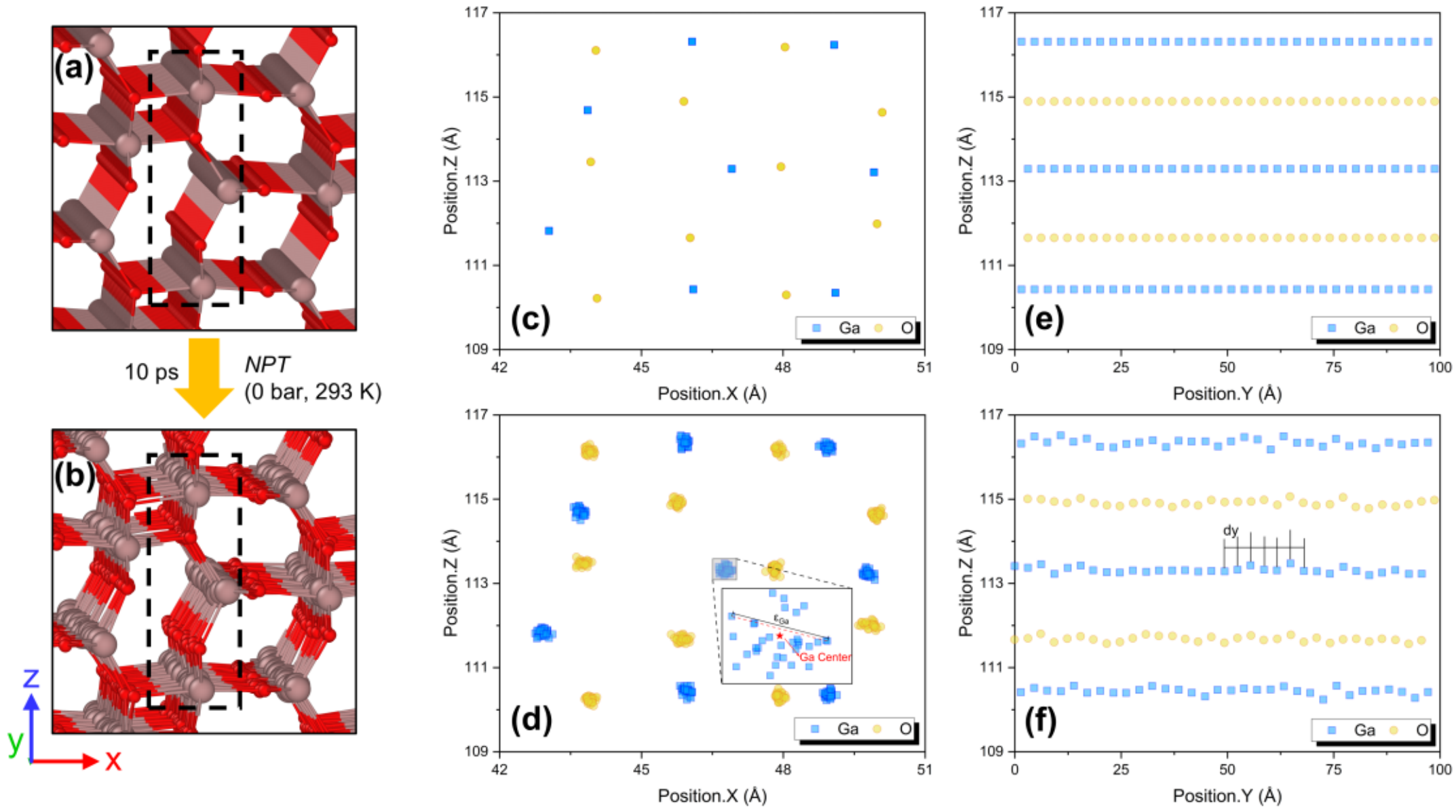

**Figure S3.** (a, b) Local structure of $\beta$-$Ga_2O_3$ before and after relaxation; (c, d) Atomic projections on the $xz$-plane before and after relaxation; (e, f) Atomic projections on the $yz$-plane before and after relaxation (region marked by the dashed box in (a, b)).

## Appendix B: Assessing the robustness of the DBSCAN identification method to thermal vibrations and high defect densities.

To evaluate the robustness of the DBSCAN method against thermal disturbances, as shown in Figure S4, we compared the identification results of the WS and DBSCAN algorithms after 10,000 relaxation steps following single-ion implantation at three different temperatures. At a lower temperature (293 K), the WS method misidentified interstitial atoms as lattice atoms due to unit-cell partitioning. In contrast, at a higher temperature (873 K), pronounced lattice thermal vibrations frequently caused the WS method to misclassify lattice atoms as interstitials. In comparison, the DBSCAN method determines the clustering radius according to the specific relaxation temperature and system, enabling more effective differentiation between lattice atoms and interstitial atoms within the $\beta$-$Ga_2O_3$ crystal.

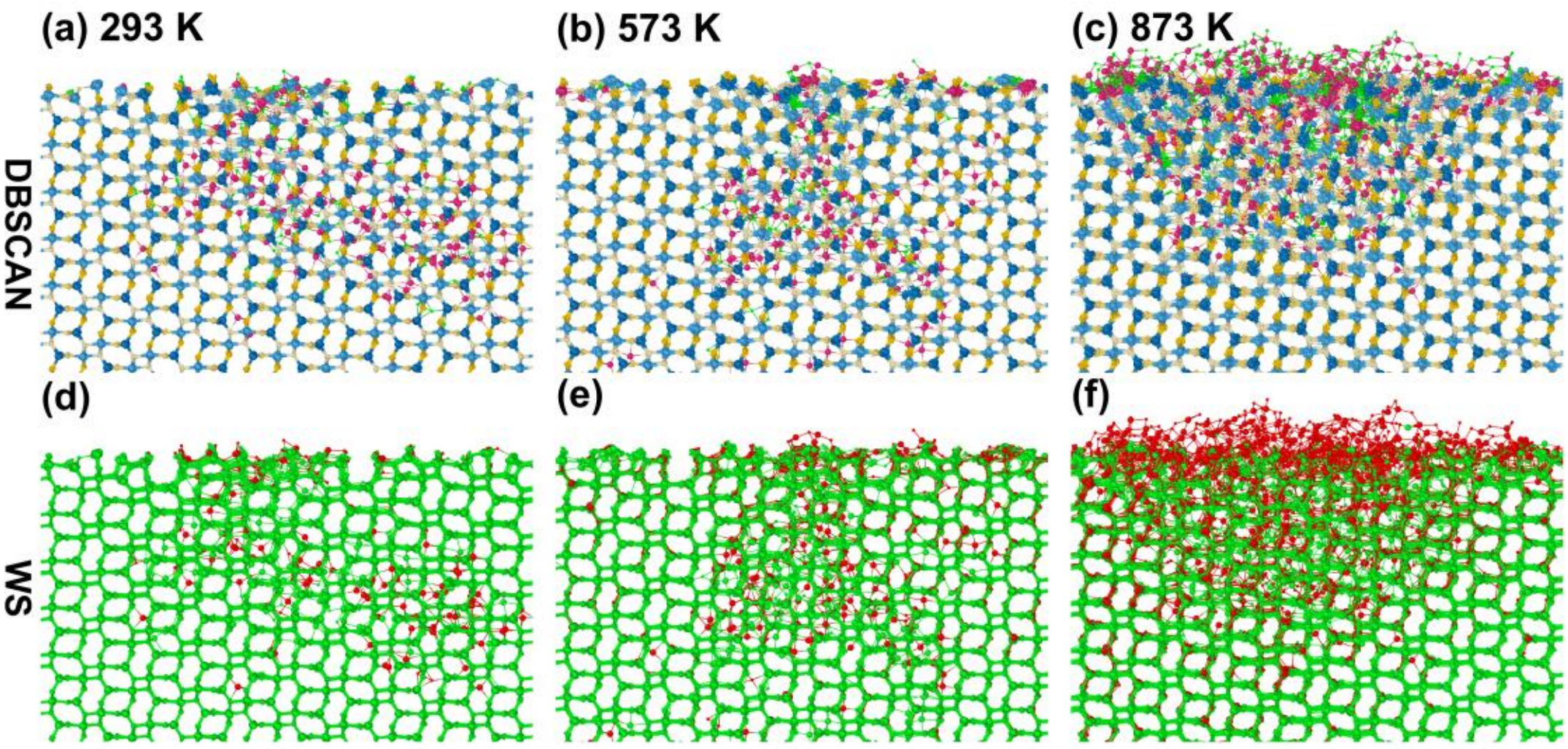

**Figure S4.** Comparison of detection results between the DBSCAN and WS methods after single-ion implantation at different temperatures. In the DBSCAN results (a–c), pink and green atoms represent $Ga_i$ and $O_i$, respectively, while the remaining atoms correspond to lattice sites. In the WS results (d–f), red atoms denote interstitials and green atoms represent lattice sites.

To evaluate the performance of the DBSCAN method under high-density defect conditions, as shown in Figure S5, we present the detection results after implantation at three different fluences (1, 3, 5 × $10^{14}$ $cm^{-2}$). To assess the effective discrimination between interstitial and lattice atoms, interstitial atoms were subsequently removed. The results show that the crystal structure of $\beta$-$Ga_2O_3$ on the (010) plane retains a high degree of integrity, with only a small number of lattice sites missing. These missing sites correspond to atomic columns that were completely damaged (i.e., contained fewer atoms than the minimum cluster-size threshold).

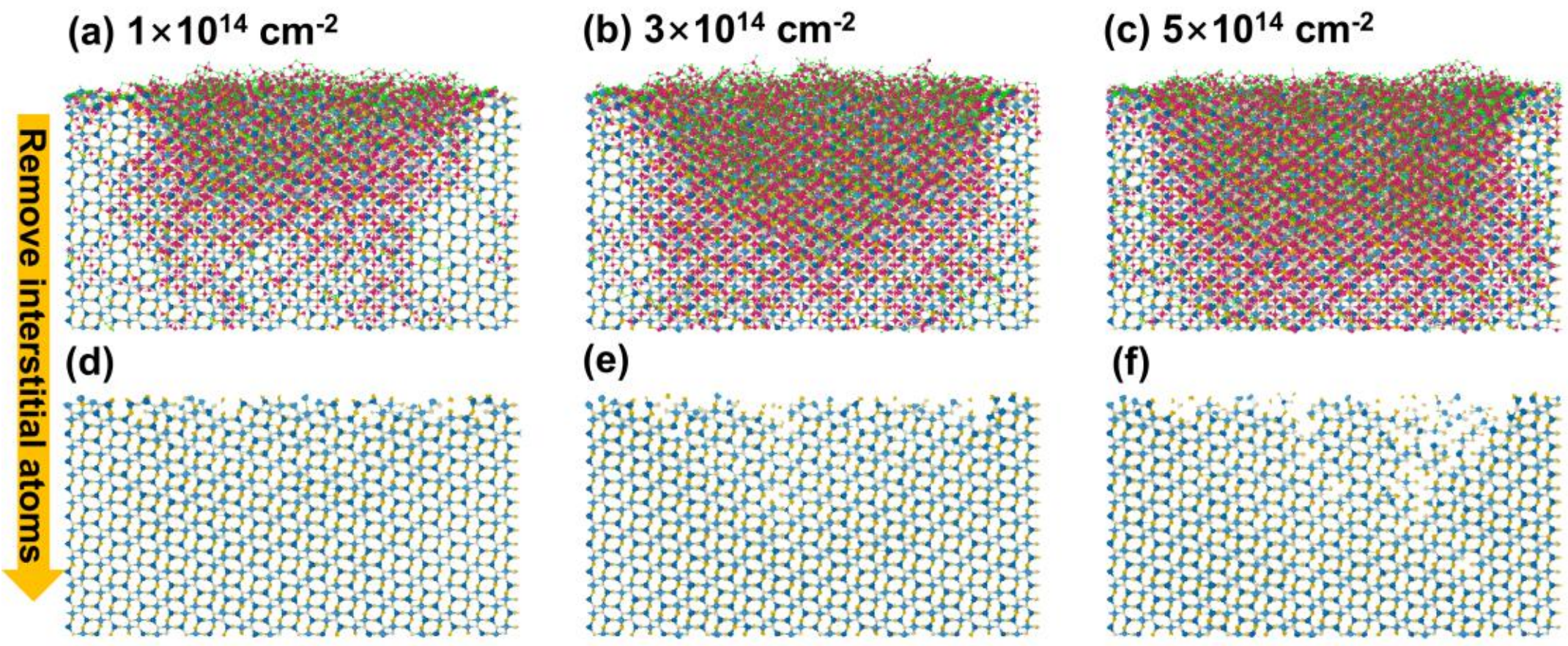

**Figure S5.** Detection results of the DBSCAN method at different implantation fluences. (a–c) Interstitial atom distributions at varying implantation fluences. (d–f) The remaining lattice after removal of interstitials.

## Appendix C: Atomic configurations of eight $Ga_i$ sites

Figure S6 illustrates eight commonly observed $Ga_i$ sites identified during the simulations. Among these, $Ga_{ia}$, $Ga_{id}$, $Ga_{ie}$, $Ga_{ig}$, and $Ga_{ih}$ are located at tetrahedral centers, whereas $Ga_{ib}$, $Ga_{ic}$, and $Ga_{if}$ reside at octahedral centers. As indicated by the dashed boxes, these interstitial atoms frequently form composite structures with adjacent vacancies. When bonds are created using a cutoff radius of 2.5 Å, these $Ga_i$ spontaneously form connections with nearby $V_{Ga}$ (these atom–vacancy links represent composite configurations and are not actual chemical bonds; to avoid confusion, bonds between $Ga_i$ and $V_{Ga}$ are shown only when the visualization is restricted to $Ga_i$ and $V_{Ga}$). Among them, $Ga_{id}$, $Ga_{ie}$, $Ga_{ig}$, and $Ga_{ih}$ often adopt *y*-shaped composite arrangements with neighboring vacancies, while the remaining $Ga_i$ sites typically form $V_{Ga}$-$Ga_i$-$V_{Ga}$ split-vacancy.

To improve the identification accuracy of the eight $Ga_i$ sites, after analyzing the coordination of each $Ga_i$ with surrounding O atoms or vacancies, the spatial distribution of the relative positions of the coordinating atoms around the $Ga_i$ center in the *xz*-plane was further examined. Taking the circular regions in Figure S6 as an example, two categories of local coordinate systems (including the mirrored cases after 180° rotation) were defined according to the type of coordination polyhedron in which the $Ga_i$ resides. A $Ga_i$ is assigned a specific interstitial type only when the number of O atoms or vacancies in each designated region satisfies the requirement for that site type (an error of one missing O atom or vacancy is allowed in regions where the expected number is two).

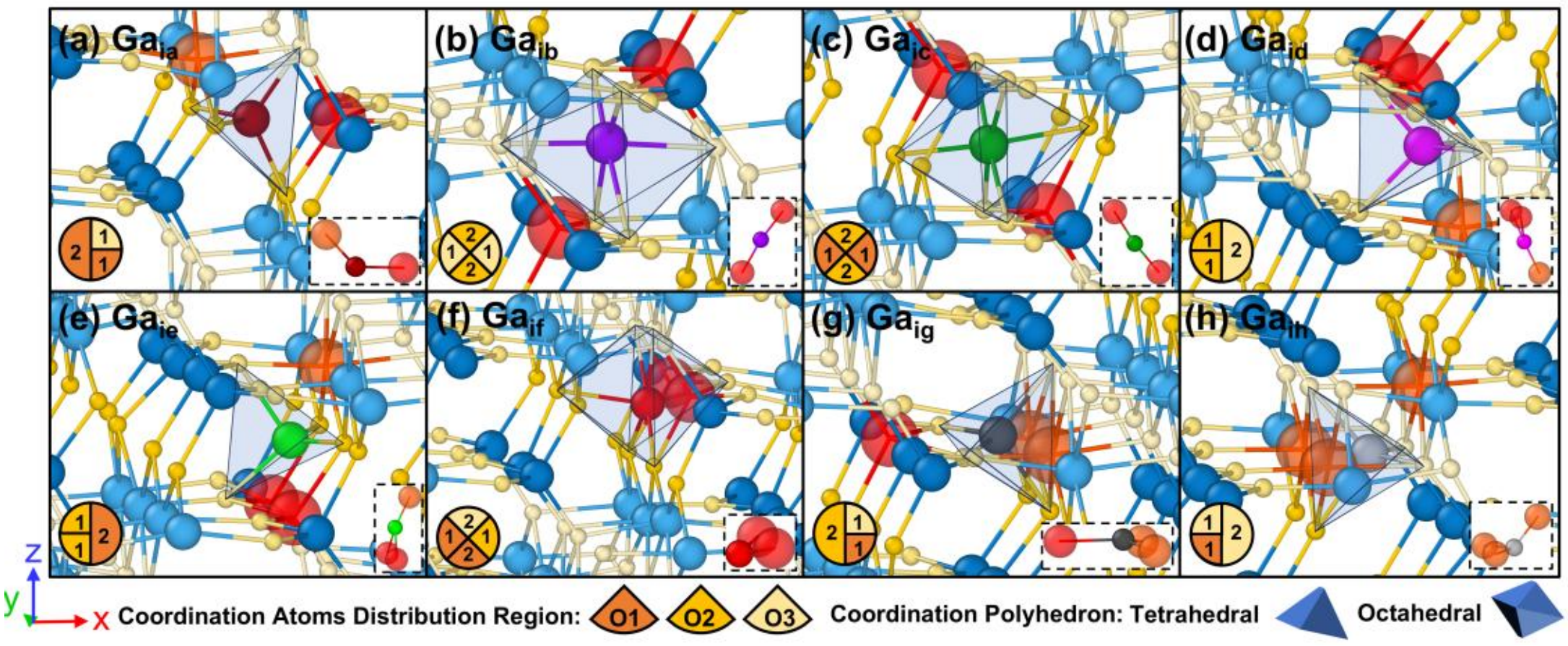

**Figure S6.** The atomic configurations of the eight $Ga_i$ sites. The circular regions illustrate the distribution and number of surrounding O atoms or vacancies in the *xz*-plane centered on each $Ga_i$ site, where the gradient from dark to light yellow represents O1, O2, and O3 atoms, respectively. The blue polyhedra denote the coordination polyhedron of each $Ga_i$ site. The dashed boxes highlight the composite structures

formed by $Ga_i$ and neighboring $V_{Ga}$.

## Appendix D: The effect of electronic stopping power on local temperature

The local temperature is computed from the atomic velocities within a 10 Å-radius volume element centered on a given atom, as given in formula (S1):

$$T_i = \frac{1}{DimN_iK_B}\left[\sum_{i=1}^{N_i}(\mathrm{m}_i\mathrm{v}_i^2)\right], \tag{S1}$$

where, $m_i$ represents the mass of a volume element, $v_i$ denotes its velocity, $T_i$ is the temperature of the $i$-th volume element, $Dim$ is the simulation dimensionality (for a three-dimensional system, the value of $Dim$ is 3), and $N_i$ is the number of atoms within the $i$-th volume element, $K_B$ is the Boltzmann constant.

Temperature analysis was conducted for single-ion implantation events at 0 K. Upon interaction with the substrate, the implanted ion transfers substantial kinetic energy to the substrate, resulting in localized high temperature near the implantation site. After approximately 2.5 ps, the substrate entered a heat dissipation phase. As shown in Figure S7, cross-sectional (010) views at 2.5, 5, 7.5, and 10 ps after implantation were analyzed. Heat diffuses radially outward from the source. Over time, the high-temperature region (red area in the Figure S7) was observed to evolve differently under two conditions: with and without electronic stopping. Comparatively, when electronic stopping was included, the high-temperature region was significantly smaller than that when electronic stopping was omitted. Moreover, the high-temperature region shrank more rapidly over time when electronic stopping was taken into account, indicating that neglecting electronic stopping leads to an overestimation of temperature.

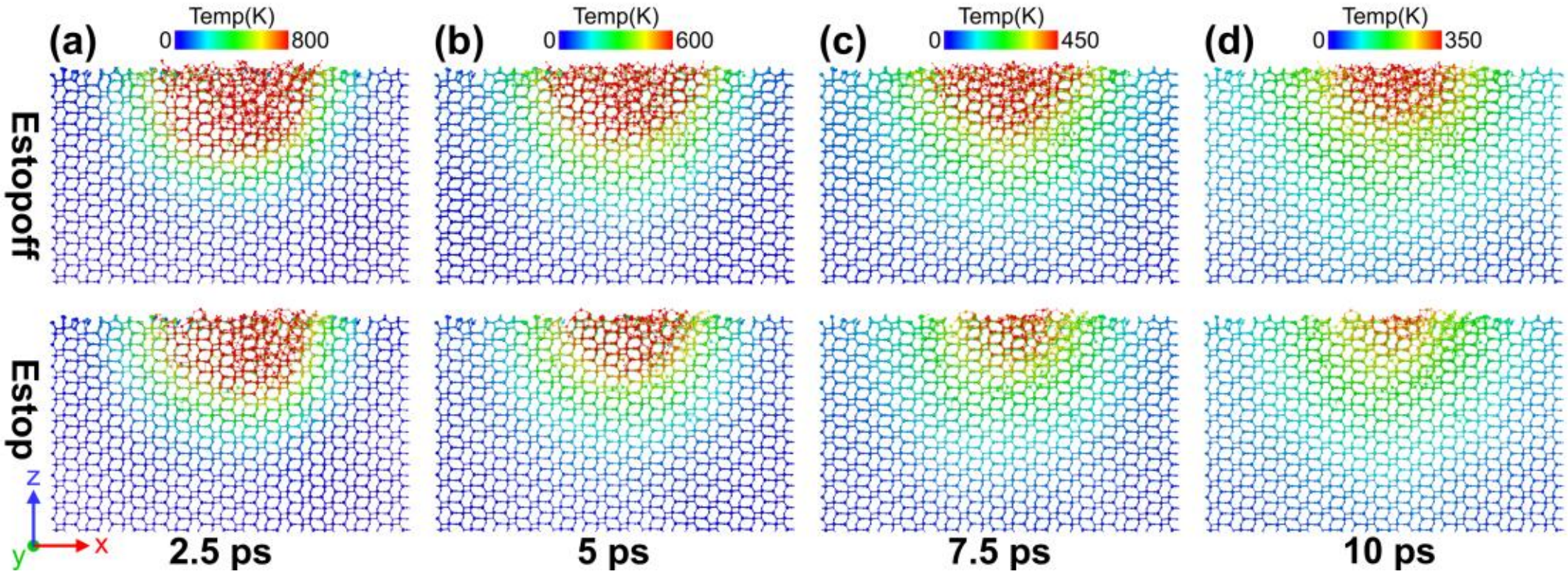

**Figure S7.** Cross-sectional (010) views of the temperature contour maps at (a) 2.5 ps, (b) 5 ps, (c) 7.5 ps, and (d) 10 ps after implantation. The first row and the second row correspond to the cases without and with electronic stopping, respectively. The color bar represents the temperature scale.

## Appendix E: Vacancy evolution and distribution analysis

As shown in Figure S8, the vacancies within the crystal after each ion implantation and relaxation cycle were statistically analyzed. The evolution of vacancies, similar to that of interstitials, also increases with the ion fluence. For $V_{Ga}$, the $V_{Ga1}$ located at the tetrahedral center is significantly more prone to formation than $V_{Ga2}$ at the octahedral center. As for $V_O$, $V_{O1}$, $V_{O2}$, and $V_{O3}$ exhibit nearly identical evolutionary trends.

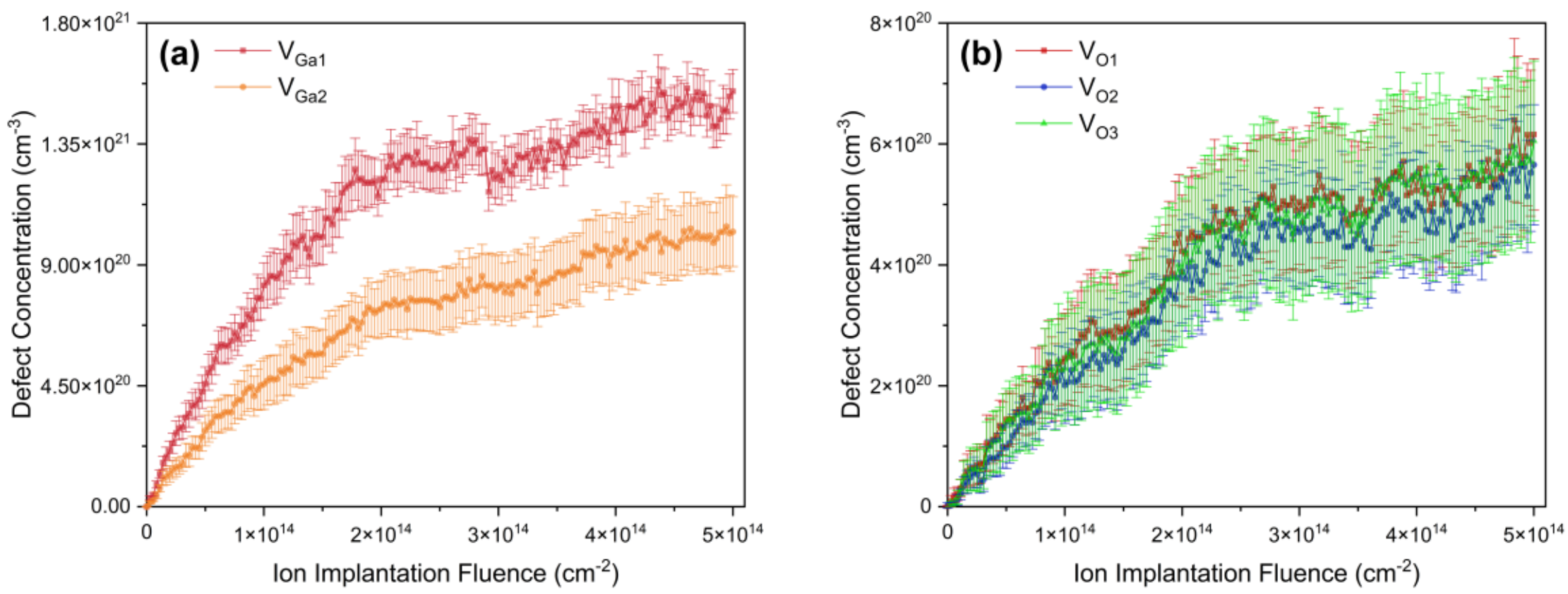

**Figure S8.** (a) Evolution of $V_{Ga1}$ and $V_{Ga2}$ as a function of implantation fluence. (b) Evolution of $V_{O1}$, $V_{O2}$ and $V_{O3}$ as a function of implantation fluence. Error bars show standard deviations.

Following the analysis of vacancy concentration profiles, we examined the distributions after annealing at three different temperatures, as shown in Figure S9. For $V_{Ga}$, a higher annealing temperature resulted in a lower concentration of $V_{Ga}$ within the crystal. Regarding $V_O$, annealing at 1237 K and 1373 K yielded comparable effects, whereas annealing at 1473 K may disrupt the O-sublattice, leading to an increase in $V_O$ concentration. Additionally, annealing at 1473 K induced surface amorphization, evidenced by an increased population of interstitial atoms near the surface. Therefore, 1373 K was selected as the optimal annealing temperature.

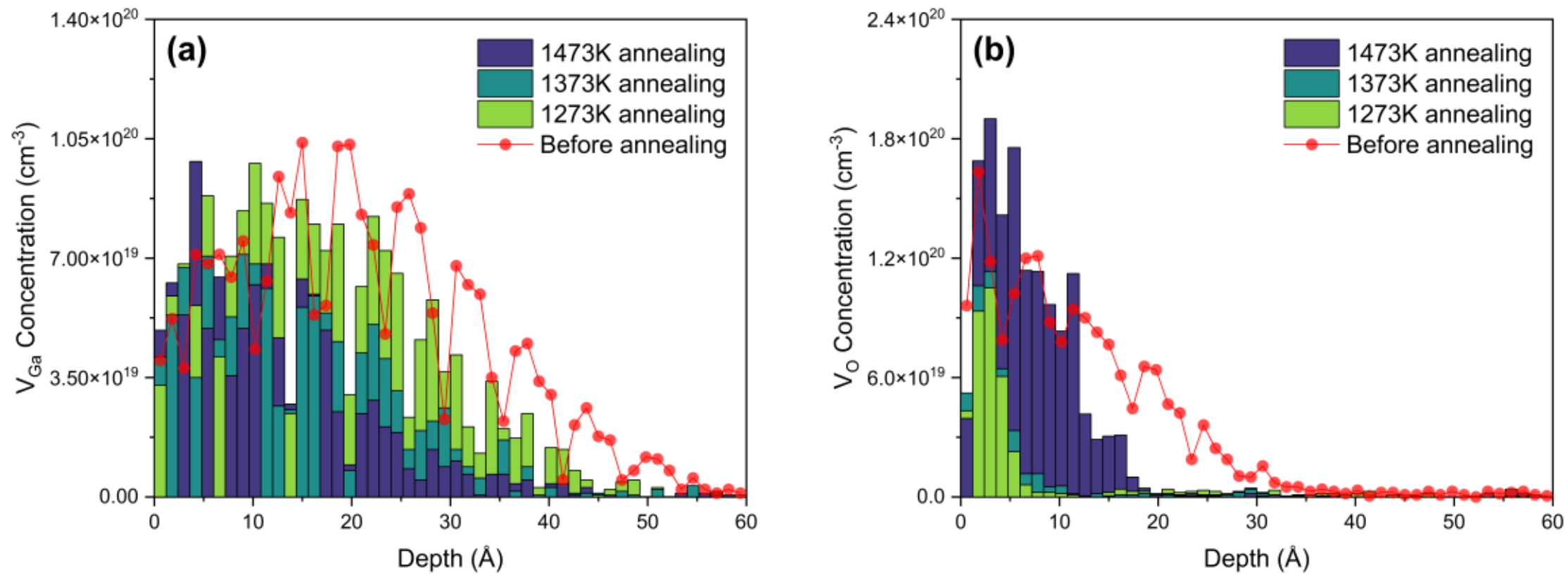

**Figure S9.** Annealing temperature test at $2 \times 10^{14}$ cm$^{-2}$. (a) $V_{Ga}$ concentration profile before and after annealing. (b) $V_O$ concentration profile before and after annealing.

Finally, as shown in Figure S10, we present the vacancy concentration profiles after implantation and subsequent annealing at different fluences. The distribution of $V_{Ga}$ extends to greater depths and covers a broader region within the crystal, whereas $V_O$ tends to accumulate closer to the surface. The aggregation of $V_{Ga}$ exhibits multiple peak-like features, similar to the distribution of Ga atoms in a perfect model, indicating that the Ga-sublattice sustains relatively minor damage upon implantation. In contrast, $V_O$ aggregation lacks such distinct peaks, suggesting that the O-sublattice tends to become amorphized after implantation. After annealing, the peak-like distribution of $V_{Ga}$ becomes more pronounced, implying a certain degree of recovery in the Ga-sublattice near the surface, although a considerable amount of $V_{Ga}$ still remains. Meanwhile, the concentration of $V_O$ decreases significantly after annealing, further demonstrating the relatively high rigidity of the O-sublattice, which facilitates recrystallization during the annealing process.

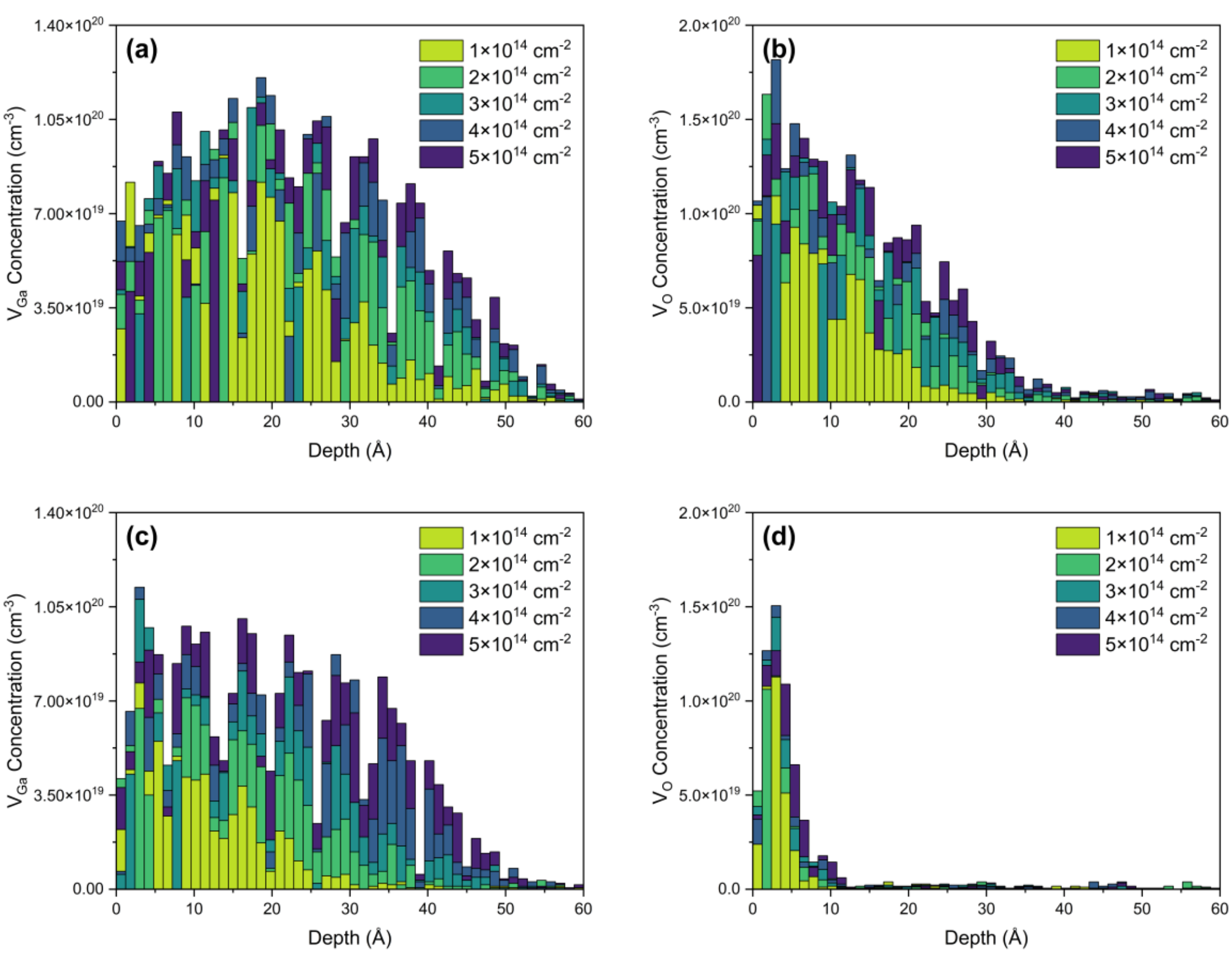

**Figure S10.** (a, b) As-implanted state and (c, d) after annealing at 1373 K. (a, c) $V_{Ga}$ concentration profile. (b, d) $V_O$ concentration profile.

## Appendix F: Lattice damage and phase transformation

To evaluate lattice damage and phase transitions, we analyzed a high-defect region (located at a depth of 10–50 Å) of the same size as the implanted area ($6 \times 6$ nm$^2$, which was uniformly damaged by ion implantation). The lattice damage was assessed using DPA (calculated as the ratio of interstitials to all atoms, and for sublattice DPA, the calculation was restricted to the sublattice). Phase transitions were evaluated using the method of Azarov *et al*. [2], in which the similarity between the Ga–Ga PRDF in the range of 4–5.2 Å and those of the standard $\beta$- and $\gamma$-phases was quantified via the Pearson correlation coefficient (Pr). The calculation of Pr follows formula (S2):

$$Pr = \frac{\sum_{i=1}^{n} (X_i - X)(Y_i - Y)}{\sqrt{\sum_{i=1}^{n} (X_i - \bar{X})^2 \sum_{i=1}^{n} (Y_i - \bar{Y})^2}}, \quad \text{(S2)}$$

where, *Xi and Yi* represent the values of the two datasets being compared (e.g., the calculated Ga–Ga PRDF and the reference PRDF of a standard $\beta$- or $\gamma$-phase) at the $i$-th radial distance, $\bar{X}$ and $\bar{Y}$ denote their respective mean values, and $n$ is the total number of data points within the evaluated radial range (e.g., 4–5.2 Å).

Figure S11(a) shows the damage levels identified by the DBSCAN and WS methods during the ion implantation stage. The damage level increases with increasing fluence. Due to the highly symmetric face-centered cubic (FCC) arrangement of O atoms in the O-sublattice, the WS method performs well for such an ordered structure, yielding results similar to those of DBSCAN. In contrast, for the Ga-sublattice with its low-symmetry in atomic arrangement, the damage levels identified by the WS method are generally lower than those obtained with DBSCAN. This leads to an overall underestimation of lattice damage in $\beta$-$Ga_2O_3$. Figure S11(b) displays the phase transformation of the crystal. As the fluence increases, the similarity of the Ga–Ga PRDF to that of the $\beta$ phase decreases, while its similarity to the $\gamma$ phase increases. The phase transition threshold occurs within the fluence range of approximately 1.65 to $2 \times 10^{14}$ cm$^{-2}$ (light gray shaded area), which corresponds to a damage level of about 0.25–0.3 dpa as identified by the DBSCAN method.

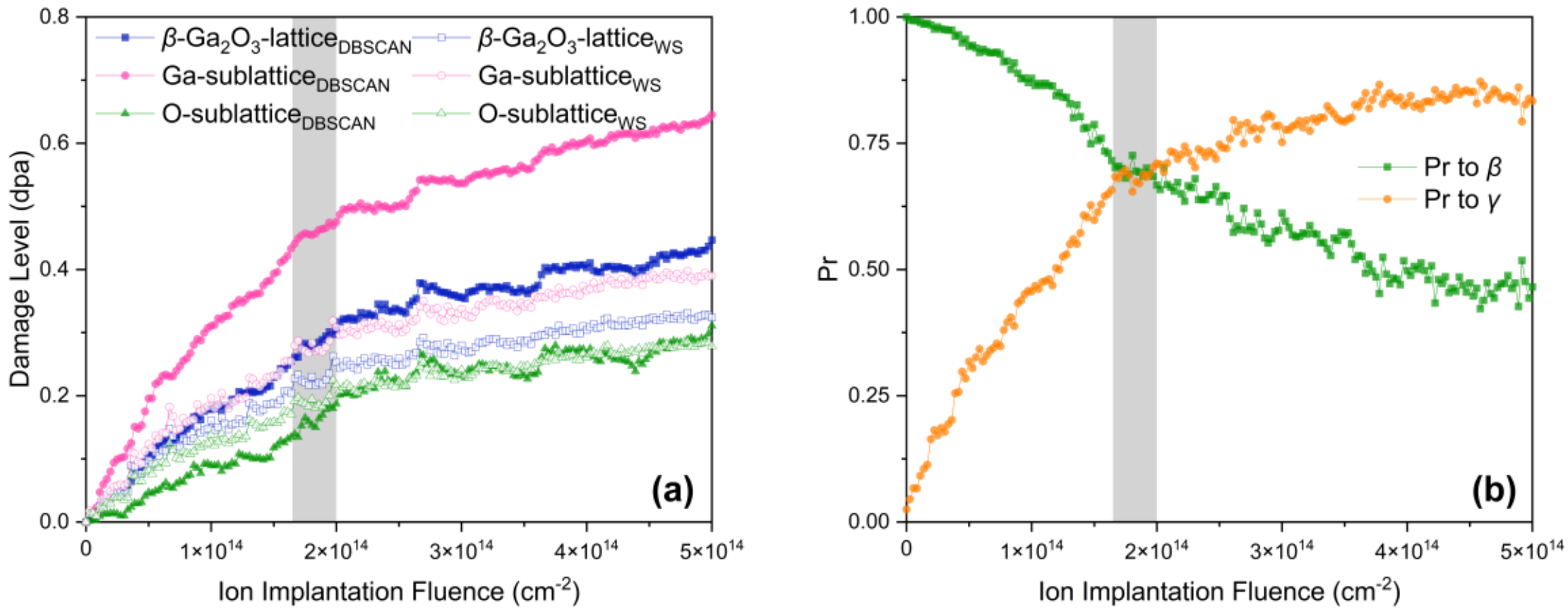

**Figure S11.** (a) Evolution of damage levels identified by DBSCAN and WS methods as a function of ion implantation fluence. (b) Evolution of Ga–Ga PRDF in the radial range 4–5.2 Å and the Pr with respect to the perfect *β*- and *γ*-phases as a function of ion implantation fluence. The light-gray shaded area indicates the phase-transition threshold region (1.65 to $2 \times 10^{14}$ $cm^{-2}$).

Next, we evaluated the damage and phase transformation of the crystals after ion implantation with five fluence levels (1 to $5 \times 10^{14}$ $cm^{-2}$) followed by annealing at 1373 K, as shown in the first rows of Figures S12(a) and (b). Below a fluence of $2 \times 10^{14}$ $cm^{-2}$, damage accumulated rapidly. Above this fluence—corresponding to the phase transformation threshold discussed in Figure S11—the fraction of the *γ*-phase Pr exceeded that of the *β*-phase, the damage accumulation rate slowed, and the tendency for phase transformation increased. The second rows of Figures S12(a) and (b) reveal that after annealing, the O-sublattice recovered significantly, while the Ga-sublattice showed only partial recovery; the residual damage level still increased with fluence. As indicated by the Pr values, annealing mainly promoted the recovery of the *β*-phase at fluences below $4 \times 10^{14}$ $cm^{-2}$, whereas above this fluence, the extent of transformation to the *γ*-phase surpassed the lattice recovery toward the *β*-phase. The irreversible phase transformation threshold after annealing is approximately 0.4 dpa, which is lower than the experimental value (0.65–0.85 dpa). This discrepancy can be attributed to factors such as defect recombination and repetitive atom displacement in actual material systems, which consume part of the damage and thus require a higher disorder level to achieve the observed transformation [3].

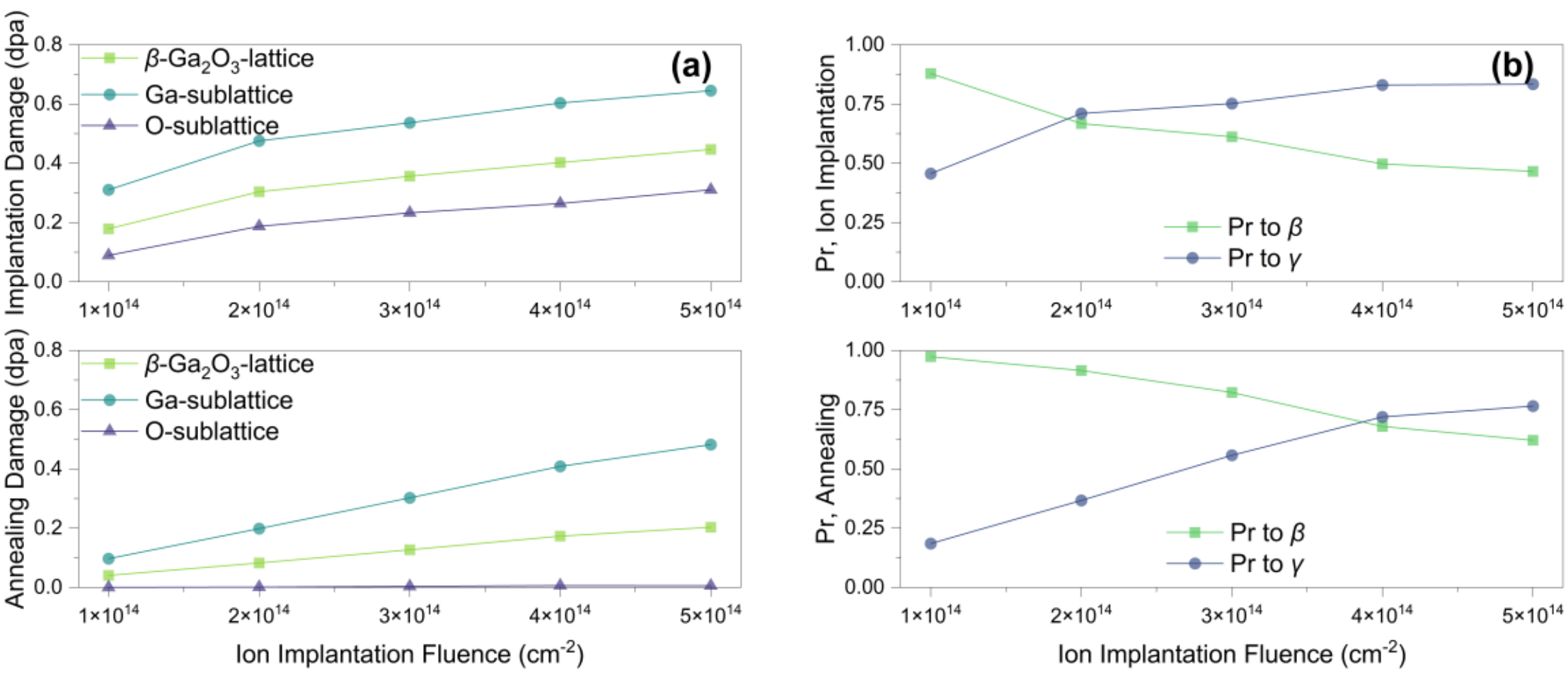

**Figure S12.** (a) Damage level and (b) Ga–Ga PRDF versus the Pr of the perfect $\gamma$- and $\beta$-phases under five ion implantation fluences (1 to 5 × $10^{14}$ $cm^{-2}$). The first row corresponds to the as-implanted state, and the second row to the stage after annealing at 1373 K.

**Appendix G: Defect comparison before and after annealing under different ion implantation fluences.**

As shown in Figure S13, lattice damage is generated during the implantation stage, with $Ga_i$ and $O_i$ arranged in a disordered state. The defective region expands as the implantation fluence increases, and the surface topological structure exhibits slight swelling. After 1373K annealing, the majority of $O_i$ within the bulk vanish, with most remaining $O_i$ distributed near the surface. The primary defects retained internally are $Ga_i$, which are arranged in an ordered configuration. Moreover, the area of this ordered structure increases with higher implantation fluences.

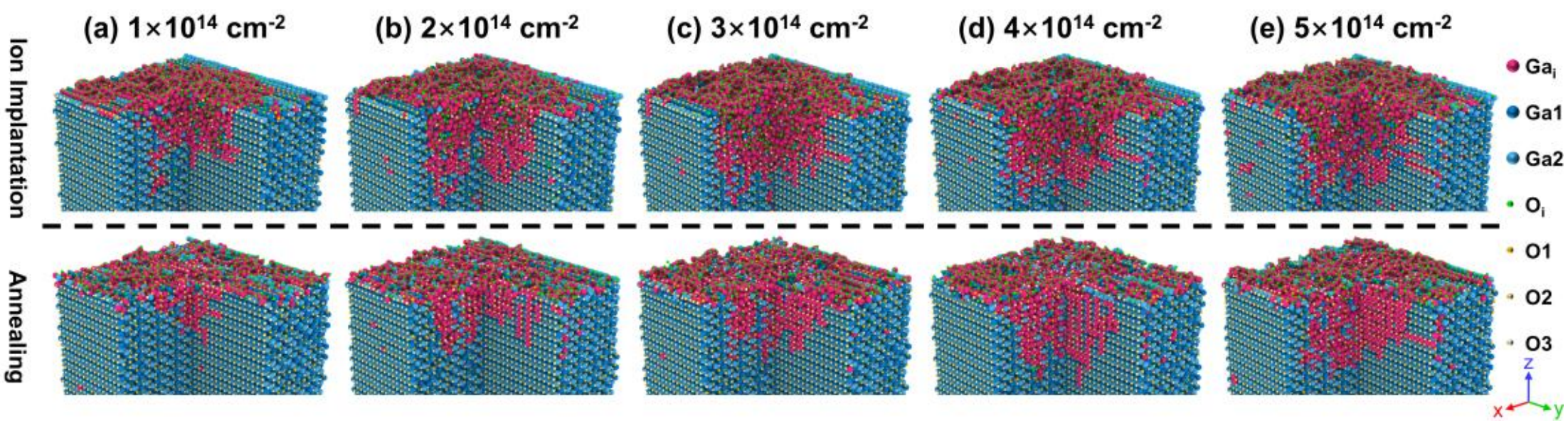

**Figure. S13.** Three-dimensional distributions of interstitials after ion implantation and subsequent annealing under five different fluences ((a–e) correspond to 1 to 5 × $10^{14}$ $cm^{-2}$, respectively). The first row shows the as-implanted state, and the second row shows the state after annealing at 1373 K.

To further investigate the point defects in $\beta$-$Ga_2O_3$, the ratios of point defect quantities after annealing to those before annealing were analyzed, as shown in Figure S14. The results reveal that $V_{Ga1}$ is retained more readily than $V_{Ga2}$ after annealing, indirectly indicating that interstitial atoms migrate more easily to the $Ga_2$ lattice site at the octahedral center compared to the $Ga_1$ site at the tetrahedral center. Meanwhile, the proportion of $V_O$ decreases significantly, demonstrating that annealing substantially promotes the recrystallization of the O-sublattice. Moreover, the proportions of $V_{Ga}$ and the eight interstitial $Ga_i$ sites (from $Ga_{ia}$ to $Ga_{ih}$)—whose ratios even exceed the pre-annealing levels—also increase with higher implantation fluences followed by annealing. This suggests that the accumulation of these point defects is associated with the phase transition from $\beta$- to $\gamma$-$Ga_2O_3$, where defect accumulation likely drives the phase transformation. It also indicates that the phase change becomes irreversible after annealing under high-fluence implantation.

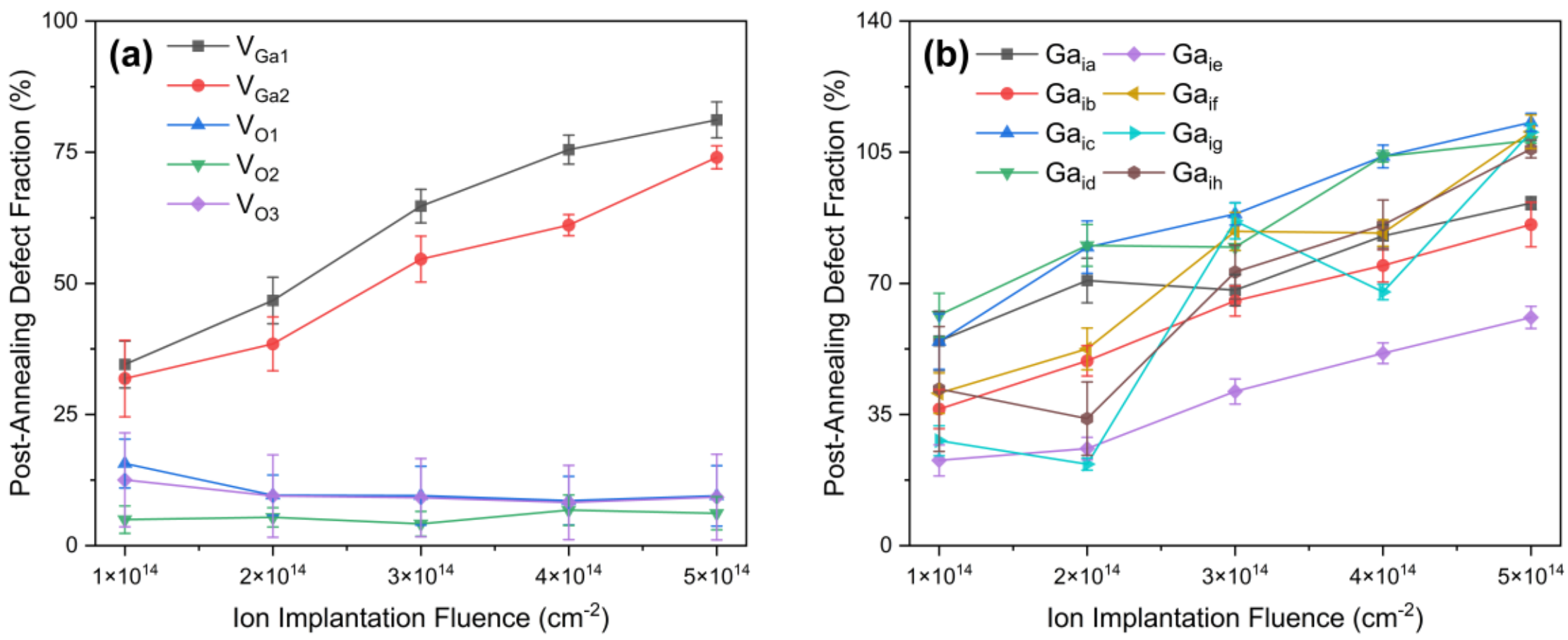

**Figure S14.** The fraction of point defects after annealing at different fluences relative to that before annealing: (a) vacancies; (b) eight $Ga_i$ sites: from $Ga_{ia}$ to $Ga_{ih}$. Error bars show standard deviations.

**Appendix H: Distribution of hydrostatic stress for Ga and O during ion implantation and annealing.**

It is crucial to analyze the stress distribution and its evolution in the model during ion implantation and annealing processes. For discrete atomic systems, the virial stress is widely applied. Consequently, the stress state of an individual atom is computed using the virial stress. The Virial stress of an atom can be obtained as the superposition of a kinetic component associated with the velocities of its neighboring atoms and a potential component related to the relative positions between atoms (in this work, a volume element with a radius of 10 Å is adopted). The stress of individual atoms can be estimated using formula (S3):

$$\sigma_{\alpha\beta}(i) = \frac{1}{N}\left(\frac{m_i v_i^\alpha v_i^\beta}{V_i} + \frac{1}{2V_i}\sum f_{ij}^\alpha r_{ij}^\beta\right), \tag{S3}$$

where, $m_i$ denotes the mass of atom $i$, $V_i$ is the volume of space associated with atom $i$, and $N$ is the number of atoms within this volume. The terms $v_i^\alpha$ and $v_i^\beta$ represent the velocity components of atom $i$ in the $\alpha$- and $\beta$-directions, respectively, while $f_{ij}^\alpha$ and $r_{ij}^\beta$ are the force component on atom $i$ from atom $j$ in the $\alpha$-direction and their separation in the $\beta$-direction, respectively. After the stress tensor for each central atom is computed, the average hydrostatic stress is obtained via formula (S4), which is used to characterize the overall stress state in the model:

$$\sigma_{hydro} = \frac{\sigma_{xx} + \sigma_{yy} + \sigma_{zz}}{3}. \tag{S4}$$

As shown in Figure S15, hydrostatic stress analysis was performed for all Ga and O atoms within a depth of 50 Å. After ion implantation, Ga atoms tend to exhibit tensile stress (stress > 0), while O atoms tend to exhibit compressive stress (stress < 0). After annealing at higher fluences, Ga atoms show a shift toward higher tensile stress, suggesting that a phase transformation ($\beta$-to-$\gamma$) may occur in the Ga-sublattice after annealing. In contrast, the stress distribution of O atoms after annealing at different fluences is generally concentrated near zero stress, indicating that the O-sublattice possesses relatively high rigidity and can be easily restored after annealing.

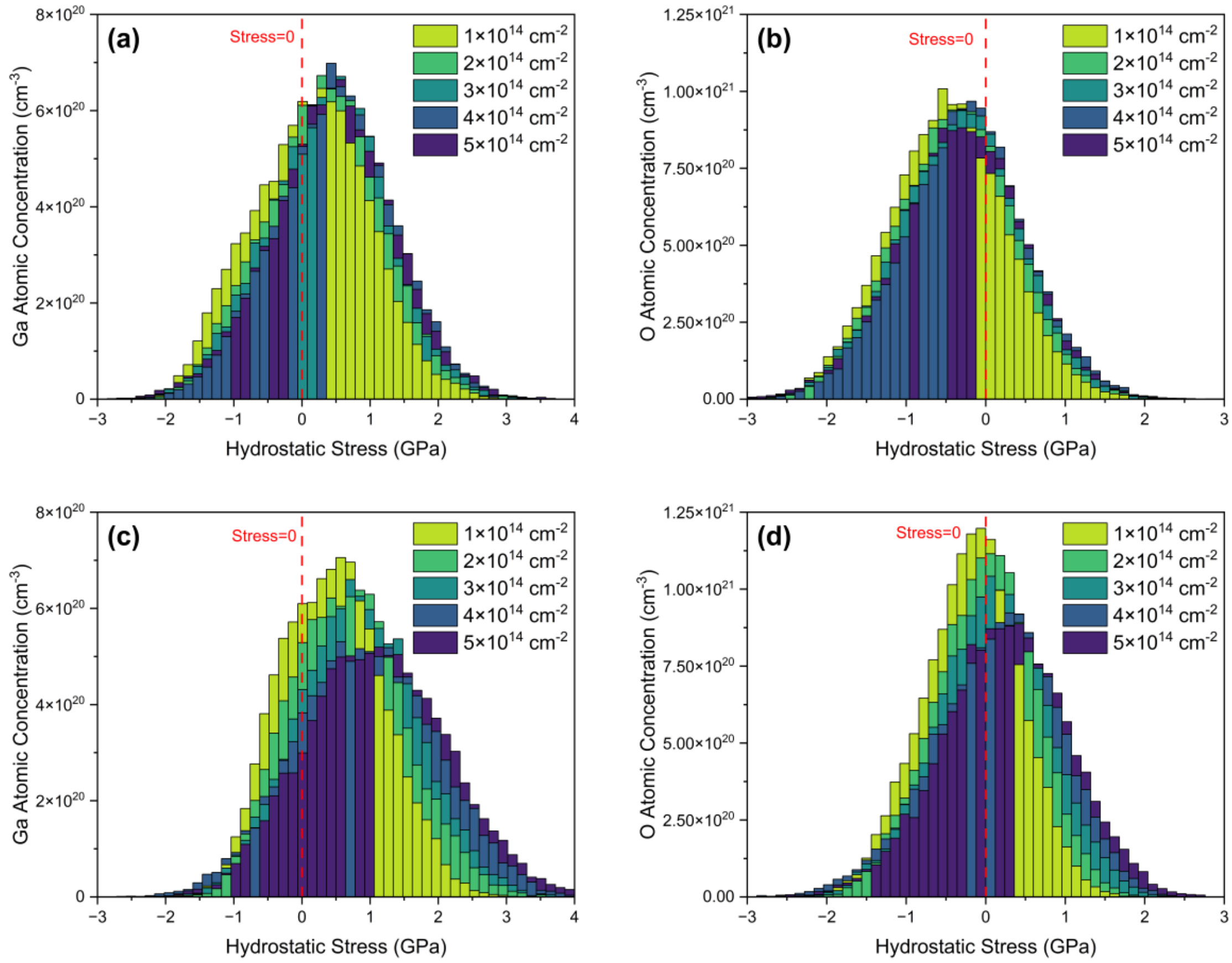

**Figure S15.** Histograms showing the distribution of atomic hydrostatic stress: (a, b) the ion implantation stage. (c, d) the annealing stage at 1373 K. (a, c) Ga stress. (b, d) O stress. The red reference line indicates zero stress.

For the $5 \times 10^{14}$ cm$^{-2}$ ion implantation model, further investigation was conducted on the effect of annealing at 1373 K on the stress in $\beta$-$Ga_2O_3$. As shown in Figure S16(a), the temperature and the pressure evolution of the system boundaries ($x$- and $y$-directions) during annealing in the *NPT* ensemble (0 bar) are presented. During the holding stage (54–1054 ps), the pressure fluctuates near 0 GPa. In contrast, larger pressure fluctuations occur during the heating (before 54 ps) and cooling (after 1054 ps) stages, which can be attributed to vigorous atomic motion and crystal-volume variations. The stress migration resulting from stable defect structures is mainly concentrated in the cooling stage. As demonstrated in video 3, which displays the stress evolution during cooling, tensile stress accumulates in regions with high defect density when the temperature drops to approximately 293 K. This phenomenon is primarily attributed to the formation of stable defect-complex structures ($Ga_{ia}$ to $Ga_{if}$) as seen in Figure S6. The association of surrounding $V_{Ga}$ with $Ga_i$ leads to an increase in the Ga–Ga interatomic distance in this region. The $Ga_i$ are in an equilibrium state "stretched" by the surrounding Ga atoms (which can also be interpreted as $Ga_i$ being "pulled" by neighboring $V_{Ga}$), thereby manifesting as tensile stress. Figure. S16(b) further illustrates this behavior by statistically analyzing the stress distribution of $Ga_i$ that migrate into interstitial sites within the $Ga_{ia}$-to-$Ga_{if}$ range after annealing (the atoms were identified by tracking their atomic IDs). Before annealing, a small fraction of $Ga_i$ already resides in the $Ga_{ia}$-to-$Ga_{if}$ region; however, most $Ga_i$ generated by ion-implantation-induced lattice damage remain at disordered positions, resulting in relatively low tensile stress. After annealing, these $Ga_i$ migrate to and stabilize in the $Ga_{ia}$-to-$Ga_{if}$ sites, causing the overall tensile stress to shift toward positive values, with an average stress increase of about 1 GPa. The behavior of $Ga_i$ further reveals the occurrence of phase transformation. The local map displays the stress distribution after annealing, showing that tensile stress is concentrated in high-defect regions, while compressive stress appears at the boundaries of these regions.

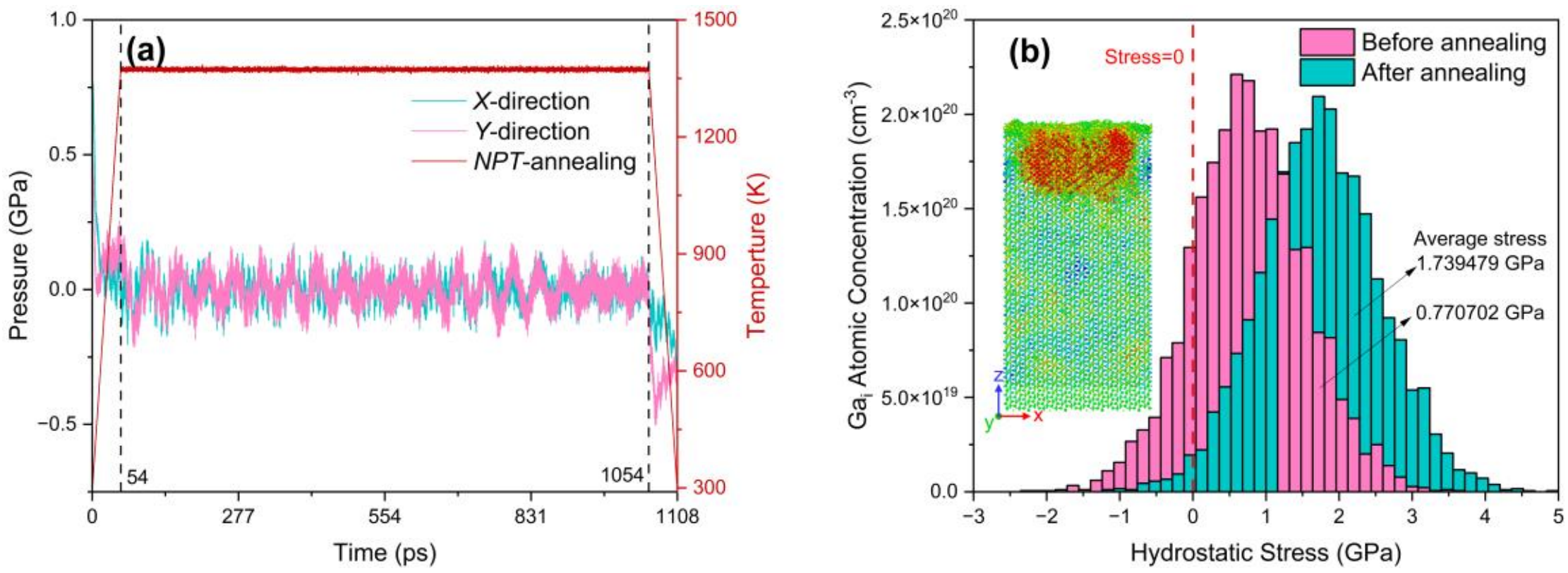

**Figure S16.** (a) Evolution of system temperature and boundary pressure for $5 \times 10^{14}$ $cm^{-2}$ fluence model annealed at 1373 K. (b) Stress distribution of $Ga_i$ (from $Ga_{ia}$ to $Ga_{if}$ after annealing) before and after annealing, with inset showing hydrostatic stress distribution contours in substrate post-annealing

## Supplementary References